\documentclass[aps,amsmath,showpacs,prd,nofootinbib]{revtex4}
\usepackage{graphicx}
\usepackage{amsmath}
\usepackage{amssymb}

\begin{document}

\title{Small amplitude quasi-breathers and oscillons}

\author{Gyula Fodor$^1$, P\'eter Forg\'acs$^{1,3}$, Zal\'an Horv\'ath$^2$,
 \'Arp\'ad Luk\'acs$^1$}
\affiliation{$^1$MTA RMKI, H-1525 Budapest 114, P.O.Box 49, Hungary,
$^2$Institute for Theoretical Physics, E\"otv\"os University,\\
 H-1117 Budapest, P\'azm\'any P\'eter s\'et\'any 1/A, Hungary,\\
{$^3$LMPT, CNRS-UMR 6083, Universit\'e de Tours, Parc de Grandmont,
37200 Tours, France}
}

\begin{abstract}
Quasi-breathers (QB) are time-periodic solutions with weak spatial localization
introduced in G.~Fodor et al. in Phys. Rev. D. {\bf 74}, 124003 (2006).
QB's provide a simple description of oscillons
(very long-living spatially localized time dependent solutions).
The small amplitude limit of QB's is worked out
in a large class of scalar theories
with a general self-interaction potential, in $D$ spatial dimensions.
It is shown that the problem of small amplitude QB's is
reduced to a universal elliptic partial differential equation.
It is also found that there is the critical dimension, $D_{\rm crit}=4$, above which
no small amplitude QB's exist.
The QB's obtained this way are shown to provide very good initial data for oscillons.
Thus these QB's provide the solution of the complicated, nonlinear time dependent problem of small
amplitude oscillons in scalar theories.

\end{abstract}

 \maketitle

\section{Introduction}\label{s:intro}
Static scalar lumps
(finite energy particle-like solutions in scalar field theories)
are known to be absent in more than two spatial dimensions ($D>2$),
however very long lived oscillating lumps
have been observed in scalar theories 
with rather general self-interaction potentials in spatial dimensions $D<7$
\cite{Kudryav,BogMak,Gleiser,CopelGM95,Gleiser04}.
These states (baptized oscillons in Ref.\ \cite{Gleiser}) are of quite some interest,
in spite of the fact that they eventually decay,
since oscillons evolve from rather generic initial data in a remarkably large class of theories.

In $2$ spatial dimensions, extremely long living breather-type objects
have been found in the sine-Gordon (sG) model \cite{PietteZakr98}, and
more recently in Ref.\ \cite{Hindmarsh-Salmi06} generic oscillons, with lifetimes
of the order of $10^6$ in natural time units have been exhibited
both in the sG and in a $\phi^{4}$ scalar model.
In $3$ spatial dimensions spherically symmetric oscillons in the
$\phi^{4}$ scalar model
with lifetimes $\approx 10^4$ have been found and investigated in detail \cite{Watkins,Honda, FFGR}.
Recently oscillons have been numerically observed starting from rather generic
initial data in  spontaneously broken SU(2) gauge theories (for the case when the mass ratio of
the Higgs field to the $W$ boson is $\approx2$), \cite{Farhi05}.
More recently 3 dimensional numerical simulations have shown \cite{Graham07} that
long living oscillons are present in the SU(2)$\times$U(1) model, i.e.\ in
the full bosonic sector of the Standard model of electroweak interactions,
which is clearly of great potential interest.
According to the findings of Ref.\ \cite{Adib}
non-symmetric oscillons evolve towards symmetric ones (at least in $1+2$ dimensions),
indicating that the long time evolutions will be dominated by spherically symmetric
configurations.

Oscillons are already interesting in their own right as very long
living lumps, appearing in many physical theories, whose existence is
due to the non-linearities.  They are expected to have important
effects on the dynamics of various systems (including the Early
Universe), since they retain a considerable amount of energy.
Importantly, oscillons can easily form in physical processes such as
the QCD phase transition, where oscillon like objects in the axion
field have been observed~\cite{Kolb:1993hw}, in vortex-antivortex
annihilation \cite{GleisThor} and in domain collapse
\cite{Hindmarsh-Salmi07}.  In~\cite{Iball} oscillating field
configurations (I-balls) were found in potentials where the quadratic
term dominates.  Once formed, oscillons could considerably influence
the dynamics of the system as has been suggested for the case of the
bubble nucleation process~\cite{Gleiser:2007ts}.  Oscillon formation
has been reported after supersymmetric hybrid
inflation~\cite{Broadhead:2005hn}. A slightly different mechanism for
the formation of long lived objects (quasi-lumps) during first order
phase transitions has been investigated in \cite{Khlopov}. The
persistence of oscillons in one spatial dimension in an expanding
background metric has been reported in Ref.\ \cite{GrahStam}.  Most
recently in a study of semiclassical decay of topological defects
possible oscillons formation has been reported~\cite{Borsanyi}.

Infinitely long lived oscillons with finite energy,
commonly known under the name of breathers, are rather exceptional.
Simple heuristic arguments indicate
that spatially localized time periodic solutions (breathers) do not exist in generic theories
\cite{Eleonski}, and that under a general perturbation  breathers are unstable
\cite{Kivshar}.
In $1$ spatial dimension
the absence of small amplitude breathers
in a scalar theory with $\phi^4$ interaction has been demonstrated in Refs.\ \cite{KS}
\cite{Vuillermot}.
More generally it has been shown in Ref.\ \cite{Kichenassamy}
that the only non-trivial $1+1$ dimensional scalar theory with real analytic self-interaction
potential which admits breather solutions is the sine-Gordon (sG) theory.
Interestingly in a recent work \cite{Arodz07}
a whole family non-radiating solutions of the $1+1$ dimensional
`signum-Gordon' model has been found. This illustrates that one can expect new and surprising
phenomena in theories where the interaction potential is not smooth.

On the other hand, bounded, time periodic solutions of nonlinear wave
equations in ${\mathbf{R}}^D$ are abundant.
By performing a Fourier decomposition in time, the problem of
finding  bounded, time periodic solutions is
reduced to solve an infinite set of coupled nonlinear partial differential
equations (PDE's). Even the simplest cases, such as $D=1$ or spherical symmetry,
when the Fourier amplitudes satisfy an infinite set of ordinary differential
equations (ODE's), are non-trivial to analyze, but it is clear that a
plethora of bounded solutions exist.
The class of bounded, time periodic
solutions in ${\mathbf{R}}^D$ contains families of breather-like objects,
which have a well defined core, outside of which the fields fall of rapidly, but barring
exceptional cases, in the far field region there is also a radiative tail, which corresponds to
a standing wave. Because of the asymptotically
standing wave asymptotics, such objects are only weakly localized in space.
The energy density of the core is much
larger than that of the tail, however, the total energy contained in
the tail is infinite. Considering such a weakly
localized object in a finite volume, $V$, its total energy, $E$, is proportional
to $E\propto V^{1/D}$.
Such objects can be thought of as radiating lumps, whose energy loss is compensated
by a flux of radiation coming from infinity, rendering the system time periodic .
In \cite{FFGR} a special class has been singled out in the huge phase space of bounded,
time periodic solutions, obtained by minimizing the energy in the
radiative tail. Solutions belonging to this class have been called as quasi-breathers (QB's).
Even if the QB's are not physical objects by themselves in the
whole space, ${\mathbf{R}}^D$, because of containing an infinite amount of energy,
nevertheless a finite piece of them (containing the core and even part of the
tail) constitutes a very good
approximation to oscillons in a ${\mathbf{R}}^D$, as demonstrated
in detail in the $\phi^4$ scalar theory in $D=3$ \cite{FFGR}.

In this paper we carry out a systematic analysis of bounded, time dependent solutions
in a $D$ dimensional scalar theory with a rather general class of self-interaction potential
in the limit when the amplitude, $\varepsilon$, of the solution goes to zero.
This way we obtain a rather general method to find small amplitude QB's.
Following Refs.\ \cite{Eleonski, KS, Kichenassamy} we derive
a formal series solution in $\varepsilon$ whose terms are all bounded both in time and space.
We show that demanding boundedness in time necessarily leads to periodicity,
and that the time and space dependence of the solution separates.
We derive a single master equation determining the spatial dependence of the leading term
in the series, which turns out to be universal for the class of scalar theories we consider.
This equation is a nonlinear elliptic PDE with a cubic nonlinearity.
It turns out that exponentially localized solutions of the master equation
exist in spatial dimension, $D<4$. While in $D=1$ the solution is unique,
for $D>1$ there is an infinite family of exponentially localized solutions.
In the case of spherical symmetry, members of this family can be characterized by the number
of their nodes. Solutions with nodes contain considerably higher energies than
the fundamental one, nevertheless they also correspond to oscillons.
The higher order terms in the small amplitude series can be obtained
from linear inhomogeneous PDE's whose source terms are
determined by the localized solution of the master equation.
Each term of the small amplitude expansion obtained this way
is exponentially localized, and one could think that it represents a
breather. In general this series
is, however, not convergent, it is rather an asymptotic one.
This fact reflects the absence of `genuine' breathers with spatial localization.
One can think of this series as an excellent approximation of QB's whose radiative
tail is smaller than any power of $\varepsilon$. Therefore the small amplitude series
provides only the exponentially localized core part of the QB. Since the
amplitude of the radiative tail is so small this does not really matter
and in fact the oscillon states corresponding
to such initial data for small values of $\varepsilon$ do have very long lifetimes.
Even the first few terms in the small amplitude expansion yield
quite good initial data for long living oscillon states, as demonstrated
by our numerical simulations for the standard $\phi^4$ theory
for the spherically symmetric case in $D=2$ and $D=3$.
As already mentioned, one of the interests of the small amplitude QB's, is that
they can be identified with the core part of small amplitude oscillons,
which radiate very weakly, and hence they have a very long lifetime.
We have verified this by numerical simulations, namely we shown
in dimensions $D=1\,,2\,,3$ that small amplitude QB's do provide
excellent initial data for long lived oscillons.

We have also computed the energy of the QB's approximated by the small amplitude series,
which corresponds to the energy content of their core and this is of course always finite.
We have shown that the energy of the QB core is a monotonously decreasing
function of the frequency near the mass threshold in dimensions $D\leq2$.
This implies the absence of a critical frequency minimizing the
energy in dimensions $D\leq2$, in agreement with known numerical results.
For $D>2$ the energy of the small amplitude QB's increases without bound
as their frequency approaches the mass threshold. This implies the existence of a critical
frequency where the energy is minimized.

As it has been already mentioned, well (exponentially) localized QB's with a finite
energy core exist only for $D<4$.
In higher dimensions, $D\geq4$ the exponential localization property of the core of
the QB's is lost and in fact the very existence of a well defined core is problematic.
In Ref.\ \cite{Gleiser04}
it has been already suggested that oscillons cease to exist,
for dimensions greater than 5 or 6 depending on the details of the potential.
According to the findings of Ref.\ \cite{SafTra} the lifetimes of
oscillons decreases rapidly as the dimensionality of space is increased,
and the QB picture does not give a good description.
We have pushed further the analysis of small amplitude oscillons
in higher dimensions to understand the situation better.
We shall present our results
in dimensions $D\geq4$ in a sequel to this paper \cite{FFHL2}.
Without going into too much details we can state the following.
Small amplitude oscillons do exist in dimensions $D\geq4$,
without any apparent limitation for $D$.
These small amplitude oscillons in dimensions $D\geq4$ are,
however, qualitatively different from their lower dimensional counterparts.
In particular, they are {\sl not well (exponentially) localized},
and they cannot be described by QB's in the sense of Ref.\ \cite{FFGR}.
The energy of higher dimensional small amplitude oscillons
also becomes quite large, therefore they are probably less interesting
physically than genuine QB's, however, by choosing
suitable initial data, they can still have very long lifetimes.
Their large energy content
may explain that such oscillons in $D>4$ are somewhat more difficult to be
found.

The plan of our paper is the following: In Section \ref{expansion} first the
class of scalar models to be studied is introduced, then the small
amplitude expansion in $D$ spatial
dimension is carried out. We derive the master equation
without any symmetry assumptions, and calculate the QB solution to order $4$ in the
$\varepsilon$ expansion as well. The energy of the QB's is also computed, and in the last
subsection the existence of the critical dimension $D=4$ is derived
above which no small amplitude QB's exist.
Section \ref{spherical} is devoted to a detailed numerical analysis
of the solutions of the spherically symmetric master equation and the explicit computations
of the higher order terms in the $\varepsilon$ expansion in spatial dimensions $D=2\,,3$.
In Section \ref{s:evolution} results on the numerical time evolution
of QB initial data up $6$th order in $\varepsilon$ is presented for $D=2\,,3$.

\section{The small amplitude expansion}\label{expansion}
In this Section we carry out a detailed analysis of the
small amplitude limit of QB's of the NLWE \eqref{e:evol}
in $D$ spatial dimensions, without any symmetry assumptions.
It turns out that bounded non-trivial small amplitude solutions are periodic in time,
and that the time and spatial dependence completely separates.
In subsection \ref{ss:master} we derive a universal
elliptic PDE, referred to as the master eq.\ governing the behaviour of the
solutions. Next, in subsection \ref{ss:higher}
the solution is obtained up to order $4$ in
$\varepsilon$ in a general class potentials, and up to order $6$ for theories
with a symmetric potential, such as the sine-Gordon model.

\subsection{The class of theories considered}\label{ss:scalar}
We consider a scalar theory in a $1+D$ dimensional flat Minkowski
space-time, with a general self-interaction
potential, whose action can be written as
\begin{equation}\label{action}
A=\int dt\, d^D\! x \left[\frac12(\partial_t\phi)^2-\frac12(\partial_i\phi)^2
-U(\phi)\right] \,,
\end{equation}
where $\phi$ is a real scalar field, $\partial_t=\partial/\partial t$,
$\partial_i=\partial/\partial x^i$ and $i=1,2,\ldots,D$.
The equation of motion following from \eqref{action}
is a non-linear wave equation (NLWE) which is given as
\begin{equation}\label{e:evol}
 -\phi_{,tt} + \Delta \phi = U'(\phi)=\phi +\sum\limits_{k=2}^{\infty}g_k\phi^k\,,\quad
{\rm where}\quad{\Delta}=\sum_{i=1}^{D}\frac{\partial^2}{\partial x_i^2}\,.
 \end{equation}
In Eq.\ \eqref{e:evol} the mass of the field is chosen to be $1$, and
it has been assumed that the potential, $U(\phi)$, can be written as a power
series in $\phi$, where the $g_k$ are real constants.
For the standard $\phi^4$ theory this interaction potential is simply
\begin{equation}
U(\phi)=\frac18\phi^2(\phi-2)^2 \,, \quad  U'(\phi)=\phi-\frac32\phi^2
+\frac12\phi^3 \,, \label{e:phi4th}
\end{equation}
i.e.\ $g_2=-\frac32$, $g_3=\frac12$ and $g_i=0$ for $i\geq 4$ in this
case. Note that in our previous paper, Ref.\ \cite{FFGR},
a different scaling of the $\phi^4$ potential has been used,
making the value of the mass to be $m=\sqrt{2}$ instead of the present value
$m=1$ used in this paper.
For the sine-Gordon potential $U(\phi)=1-\cos(\phi)$, we have
$g_{2i}=0$ and $g_{2i+1}=(-1)^i/(2i+1)!$.

The energy corresponding to the action (\ref{action}) can be written as
\begin{equation}\label{e:density}
E = \int d^{D}x\,{\cal E}\,, \quad
{\cal{E}} = \frac{1}{2} \left(\partial_t \phi\right)^2
+ \frac{1}{2} \left(\partial_i \phi\right)^2
+U(\phi)\,,
\end{equation}
where ${\cal{E}}$ denotes the energy density.

\subsection{Derivation of the master equation}\label{ss:master}

We are looking for small amplitude solutions, therefore we expand the
scalar field, $\phi$, in terms of a parameter $\varepsilon$ as
\begin{equation}
\phi=\sum_{k=1}^\infty\varepsilon^k \phi_n \,. \label{e:sumphi}
\end{equation}
In order to obtain non-trivial solutions of Eq.\ \eqref{e:evol}
their characteristic scale must also become
$\varepsilon$-dependent. The size of smooth configurations is
expected to increase for decreasing values of $\varepsilon$, therefore it is
natural to  introduce new spatial coordinates by the following rescaling
\begin{equation}
\zeta^i=\varepsilon x^i \,.
\end{equation}
One must also allow for the $\varepsilon$ dependence of
the time-scale of the configurations,
therefore a new time coordinate is introduced as
\begin{equation}
\tau=\omega(\varepsilon) t \,.
\end{equation}
$\omega(\varepsilon)$ is assumed to be analytic near the threshold, $\omega=1$,
and it is expanded as
\begin{equation}
\omega^2(\varepsilon)=1+\sum_{k=1}^\infty\varepsilon^k\omega_k \,.
\end{equation}
After these rescalings Eq.\ \eqref{e:evol} takes the following form
\begin{equation}\label{e:evoleps}
 -\omega^2\ddot\phi + \varepsilon^2\Delta \phi=
\phi +\sum\limits_{k=2}^{\infty}g_k\phi^k\,.
\end{equation}
In equation (\ref{e:evoleps}) and in the rest of this section an overdot
stands for the derivative with respect to the rescaled time coordinate, $\tau$,
and all spatial derivatives are taken with respect to the rescaled coordinates
$\zeta^i$.  Substituting the $\varepsilon$ expansion of the scalar field,
$\phi$, and of $\omega^2$ into \eqref{e:evoleps} the equations determining
the first three lowest order terms are:
\begin{eqnarray}
\ddot\phi_1+\phi_1&=&0 \,,\label{e:phi1}\\
\ddot\phi_2+\phi_2+g_2\phi_1^2+\omega_1\ddot\phi_1&=&0 \,,\label{e:phi2}\\
\ddot\phi_3+\phi_3+2g_2\phi_1\phi_2+g_3\phi_1^3-\ddot\phi_1-\Delta\phi_1
+\omega_1\ddot\phi_2+\omega_2\ddot\phi_1&=&0 \,.\label{e:phi3}
\end{eqnarray}
As it is clear from Eqs.\ \eqref{e:phi1}-\eqref{e:phi3} the time dependence
has been separated from the spatial one, and we have obtained a set
of harmonic oscillator equations. Now the solution of Eq.\ \eqref{e:phi1} is clearly
given by
\begin{equation}
\phi_1=p_1\cos(\tau+\alpha)\,,
\end{equation}
where $p_1$ and $\alpha$ are functions of the spatial variables
$\zeta^i$. That is, the lowest order term of the solution is just a harmonic
oscillator in time,
with frequency $\omega=1$ (note that this is the same
with respect to both time coordinates, $\tau$ and $t$ at this order). This distinguished
value of the frequency, $\omega=1$, corresponds to the threshold determined
by the mass of the scalar field.

As Eq.\ \eqref{e:phi2} is a linear inhomogeneous equation, its
solution is easily obtained:
\begin{equation}
\phi_2=p_2\cos(\tau+\alpha)+q_2\sin(\tau+\alpha)
+\frac{g_2}{6}p_1^2[\cos(2\tau+2\alpha)-3]
+\frac{\omega_1}{4} p_1\left[2\tau\sin(\tau+\alpha)+\cos(\tau+\alpha)\right]\,.
\end{equation}
Since we are looking for bounded solutions, it is necessary to impose
$\omega_1=0$. This amounts to demanding the absence of the resonance term
$\omega_1\ddot\phi_1$ in eq.\ (\ref{e:phi2}). Substituting the solutions
for $\phi_1$ and $\phi_2$ into Eq.\ \eqref{e:phi3} one obtains yet another forced
oscillator equation for the time dependence of $\phi_3$:
\begin{eqnarray}
&&\ddot\phi_3+\phi_3+(p_1\Delta\alpha+2\nabla\alpha\nabla p_1)\sin(\tau+\alpha)
-\left[\Delta p_1+\omega_2p_1+\lambda p_1^3-p_1(\nabla\alpha)^2\right]
\cos(\tau+\alpha)\nonumber\\
&&+\frac{1}{12}p_1^3(2g_2^2+3g_3)\cos(3\tau+3\alpha)
+g_2p_1\left[q_2\sin(2\tau+2\alpha)+p_2\cos(2\tau+2\alpha)+p_2\right]
=0\,,\label{e:phi3-1}
\end{eqnarray}
where we have introduced the combination
\begin{equation}
\lambda=\frac{5}{6}g_2^2-\frac{3}{4}g_3 \,,
\end{equation}
which will play an important r\^ole in the following. For the standard $\phi^4$
theory \eqref{e:phi4th}, this parameter takes the value $\lambda=3/2$.
For the sine-Gordon potential $\lambda=1/8$.

As already explained we are looking for bounded solutions, therefore it is necessary
to guarantee the absence of resonance terms also in Eq.\
\eqref{e:phi3-1}. The vanishing of the coefficient of the
$\sin(\tau+\alpha)$ terms implies
\begin{equation}
\nabla(p_1^2\nabla\alpha)=0\,,
\end{equation}
and from this equation one immediately derives the following condition
\begin{equation}
\int_\Omega\alpha\nabla(p_1^2\nabla\alpha)=
\int_{\partial\Omega}\alpha p_1^2 n\cdot\nabla\alpha
-\int_\Omega p_1^2(\nabla\alpha)^2=0 \,.
\end{equation}
Assuming that the integrand of the boundary term vanishes
sufficiently fast we conclude that $\nabla\alpha=0$.
Our assumption is quite reasonable since we are looking for bounded
solutions in time and localized in space. Therefore $\alpha$ must be a
constant which can be absorbed by a shift in the time variable. From
now on we set $\alpha=0$.
Then the vanishing of the coefficient of the resonance term
proportional to $\cos \tau$ implies
\begin{equation}\label{e:eqp1}
\Delta p_1+\omega_2p_1+\lambda p_1^3=0 \,.
\end{equation}
A necessary condition that this equation admit
exponentially localized solutions is $\omega_2<0$, which we shall assume from now on.
In this case we can set $\omega_2=-1$ by a simultaneous rescaling
of $\zeta^i$ and $p_1$.
This rescaling corresponds to choosing a different parametrization
$\varepsilon$ for a solution with a specific frequency $\omega$.

By an analytic redefinition of the expansion parameter, $\varepsilon$,
and by rescalings, all coefficients $\omega_i$ can be made to vanish for $i>2$.
This means that by a suitable transformation of the expansion parameter,
$\varepsilon$, one can always achieve that the following
relation between the frequency, $\omega$,
and the expansion parameter, $\varepsilon$, holds:
\begin{equation}
\omega=\sqrt{1-\varepsilon^2}\,.
\end{equation}
In terms of the physical time coordinate, $t$, the configuration
oscillates with frequency $\omega$. This is the physically important
frequency characterizing these periodic solutions. Apart from the leading order
behaviour, the precise choice of how $\varepsilon$ depends on $\omega$
is physically irrelevant.

After setting $\omega_2=-1$, it is easy to see (multiplying
(\ref{e:eqp1}) with $p_1$ and integrating) that $\lambda>0$ is a
necessary condition for the existence of bounded solutions vanishing
at infinity.  Assuming $\lambda>0$, by rescaling $p_1$ one obtains
\begin{equation}\label{e:Sequation}
\Delta S -S +S^3=0\,,\quad S=p_1\sqrt{\lambda} \,.
\end{equation}
This master equation constitutes an important result equation of our paper.
Quite remarkably Eq.\eqref{e:Sequation} is universal for the class of theories considered,
and the dependence on the parameters of the interaction potential enters only
through the combination $\lambda$ when reconstructing $\phi_1$.

\subsection{Higher orders in the $\varepsilon$ expansion}\label{ss:higher}

Let us now turn to the determination of some higher order terms in the $\varepsilon$
expansion. Armed with the simple form of the solution for $p_1$ it is now easy
to obtain the explicit time dependence of $\phi_3$ by
integrating Eq.\ \eqref{e:phi3-1}:
\begin{equation}
\phi_3=q_3\sin \tau+ p_3\cos \tau+\frac{p_1}{3}\left\{\frac{1}{8}
\left(\frac{4}{3}g_2^2-\lambda\right)p_1^2 \cos(3\tau)
+g_2 \left[q_2\sin(2\tau)+p_2(\cos(2\tau)-3)\right]\right\} \,. \label{e:phi3-2}
\end{equation}
In $\phi_3$ two new functions, $p_3$, $q_3$ have appeared.
The absence of resonances at fourth order yields two conditions:
\begin{equation}
\lambda q_2 p_1^2-q_2+\Delta q_2=0 \,, \label{e:ph4sin}
\end{equation}
which is the vanishing condition of coefficient of the term
of $\sin\tau$, and another one
\begin{equation}
3\lambda p_2p_1^2-p_2+\Delta p_2=0\,, \label{e:ph4cos}
\end{equation}
which ensures the vanishing of the coefficient of $\cos\tau$.

Let us first note, that quite remarkably $q_2\propto p_1$ actually solves Eq.
\eqref{e:ph4sin}. Therefore by shifting the time coordinate by a suitable term
of order $\varepsilon$ we can eliminate $q_2$.
In higher orders of the $\varepsilon$ expansion,
we have verified that up to order $\varepsilon^9$ the vanishing of the coefficient
of the terms proportional to $\sin \tau$
leads to equations which are equivalent to Eq.\ \eqref{e:ph4sin}.
We conjecture that this is in fact true to all orders, i.e.\
by a suitable choice of the origin of the time coordinate $\tau$ one
eliminate all terms proportional to $\sin \tau$.
This observation is quite important because it implies that
all small amplitude QB-type solutions necessarily possess time reflection symmetry.
This is of course what one would expect based on simple physical intuition.
Let us point out here, that all long-lived oscillon configurations observed in
time evolution simulations appear to show this symmetry to a very high
degree. Of course oscillons are not {\sl exactly} time reflection symmetric because
they radiate some energy to infinity.
Moreover, time reflection symmetry has been usually implicitly
assumed when performing Fourier decomposition in order to find time
periodic states. For all these reasons it would be of interest to
find a mathematical proof of
the validity of time reflection symmetry for periodic
solutions of NLWE's.

There is no reason to expect that Eq.\ \eqref{e:ph4cos} admits bounded
solutions vanishing at infinity
apart from those corresponding to the translational and rotational symmetries,
therefore from now on we set $p_2\equiv 0$.
The vanishing of the coefficient of the $\cos \tau$ term
in the fifth order equation yields
\begin{equation}\label{e:Deltap3}
\Delta p_3-p_3+3\lambda p_1^2 p_3
+\frac{g_2^2}{9}p_1\left(17p_1^2+19(\nabla p_1)^2\right)
+\frac{p_1^5}{216}(378 g_4 g_2+36\lambda g_2^2-280g_2^4-9\lambda^2-135g_5)
=0 \,.
\end{equation}
This is a linear, inhomogeneous equation. It can be brought to a much simpler form
by introducing a new variable $Z$ instead of $p_3$:
\begin{equation}
p_3=\frac{1}{\lambda^2\sqrt{\lambda}}\left[
\left(\frac{1}{24}\lambda^2-\frac{1}{6}\lambda g_2^2+\frac{5}{8}g_5
-\frac{7}{4}g_2g_4+\frac{35}{27}g_2^4\right)Z
-\frac{1}{54}\lambda g_2^2S(32+19S^2)
\right]\,. \label{e:p3}
\end{equation}
Then Eq.\ (\ref{e:Deltap3}) takes the compact form
\begin{equation}
\Delta Z-Z+3S^2Z-S^5=0\,. \label{e:Z}
\end{equation}
For the specific example of the standard $\phi^4$ theory this relation is simply
\begin{equation}
p_3=\frac{\sqrt{2}}{3\sqrt{3}}\left(
\frac{65}{8}Z-\frac{8}{3}S-\frac{19}{12}S^3
\right)\,.
\end{equation}
Integrating the corresponding equation for $\phi_5$ a new unknown
function, $p_5$ appears, in analogy to the third order case.

To summarize, we have obtained the solution of the NLWE \eqref{e:evol}
in the small amplitude expansion up to order four. All terms have harmonic
time dependence, and the spatial part is determined by the two universal
elliptic PDE's, eqs.\ (\ref{e:Sequation},\ref{e:Z}). There is no
obstacle to continue the computation to higher orders, the general
formulae become then quite complicated of course.
The small amplitude expansion of the solution of Eq.\ \eqref{e:evoleps}
up to order four for general interaction potentials
can be written as:
\begin{eqnarray}
\phi_1&=&p_1\cos \tau\label{e:phi3b1}\\
\phi_2&=&\frac16 g_2p_1^2\left(\cos(2\tau)-3\right)\label{e:phi3b2}\\
\phi_3&=&p_3\cos \tau+\frac{1}{72}(4g_2^2-3\lambda)p_1^3\cos(3\tau)
\label{e:phi3b3}\\
\phi_4&=&
\frac{1}{360}p_1^4\left(3g_4-5g_2\lambda+5g_2^3\right)\cos(4\tau)\nonumber\\
&&-\frac{1}{72}\left(8g_2(\nabla p_1)^2-12g_4p_1^4+16g_2^3p_1^4
-24g_2p_1p_3-23g_2\lambda p_1^4-8g_2p_1^2\right)\cos(2\tau)\label{e:phi3b4}\\
&&-g_2p_1^2-g_2p_1p_3+\frac{1}{6}g_2\lambda p_1^4-g_2(\nabla p_1)^2
+\frac{31}{72}g_2^3p_1^4-\frac{3}{8}g_4p_1^4 \ .\nonumber
\end{eqnarray}

A considerable simplification occurs when the scalar self-interaction
potential, $U(\phi)$, is
symmetric around its minimum $\phi=0$. In this case $g_{2i}=0$ for all
$i=1\,,\ldots$, and all even power terms in the $\varepsilon$ expansion
vanish, i.e. $\phi_{2i}=0$ for $i=1\,,\ldots$.
Since $\phi_{2n}$ contains only terms of the form
$\cos(2k\tau)$ with $k=1\,,\ldots n$, and $\phi_{2n+1}$ contains only
terms proportional to $\cos((2k+1)\tau)\,,$ with $k=1\,,\ldots n$,
this also implies that for such
symmetric potentials no even terms in the Fourier expansion arise. In
this case $p_3$ is proportional to $Z$ and the equation determining
the function $p_5$ becomes reasonably simple, it can be written as
\begin{eqnarray}
&&\Delta p_5-p_5+3S^2p_5
+\frac{SZ}{576\sqrt{\lambda}}(3Z-5S^3)
\left(\frac{15g_5}{\lambda^2}+1\right)^2\nonumber\\
&&+\frac{S^3}{32\sqrt{\lambda}}\left[(\nabla S)^2-S^2\right]
-\frac{S^7}{576\sqrt{\lambda}}\left(\frac{315g_7}{\lambda^3}
-\frac{60g_5}{\lambda^2}+1\right)=0 \,.
\end{eqnarray}
Then
\begin{eqnarray}
&&\phi_5=p_5\cos\tau
+\frac{S^5}{1152\sqrt{\lambda}}\left(\frac{3g_5}{\lambda^2}+2\right)
\cos(5\tau)\nonumber\\
&&-\frac{S}{384\sqrt{\lambda}}\left[
\left(\frac{30g_5}{\lambda^2}+2\right)SZ+12S^2-12(\nabla S)^2
-\left(\frac{15g_5}{\lambda^2}-2\right)S^4
\right]\cos(3\tau) \,.
\end{eqnarray}
These expressions encompass for example the case of the sine-Gordon
model.  The corresponding equations for the $\phi^4$ theory, in the
case of spherical symmetry, will be listed in Section \ref{spherical}.

As we have already stressed several times,
it is by now well understood that spatially localized
breathers of the NLWE \eqref{e:evol} do not exist in ${\mathbf{R}}^D$
for general analytic potentials, even
if a general mathematical proof is known only in $D=1$.
Let us note here, that a remarkable example in $D=1$
admitting non-radiating breather-type solutions in the framework of ``$V$''-shaped
(non-differentiable) potentials
evades this theorem \cite{Arodz07}.
In the case of analytic potentials, where the theorem applies there is still
a point to be stressed.
Assuming that exponentially decreasing solutions of the master equation
exist, all higher order terms in the small amplitude expansion are also
exponentially localized, and they are periodic in time.
As we have learned from the example of the one dimensional $\phi^4$ theory \cite{KS}
the series solution in powers of $\varepsilon$ does not converge to a breather,
it is an asymptotic series.
Nevertheless, to a given order in the expansion for sufficiently
small values of $\varepsilon$
the corresponding sum yields a configuration with a spatially well localized core.
This time periodic configuration corresponds to a QB whose standing wave tail is smaller
than $\varepsilon^n$ for any $n>0$. As it will be shown in Section \ref{s:evolution}
such QB's constitute an excellent approximation to an oscillon.

\subsection{The energy}\label{ss:energy}

In this subsection we evaluate the energy of small amplitude
QB's, in $D$ dimension.
In the rescaled coordinate system, $\tau\,,\zeta$,
the energy of a configuration, Eq.\ \eqref{e:density}
can be written as
\begin{equation}
E = \frac{1}{\varepsilon^D}\int d^{D}\zeta\,{\cal E}\,, \quad{\rm where}\quad
{\cal{E}} = \frac{1}{2} (1-\varepsilon^2)\left(\partial_\tau \phi\right)^2
+ \varepsilon^2\frac{1}{2} \left(\partial_i \phi\right)^2
+U(\phi)\,.
\end{equation}
Because of the periodic time dependence we shall compute
the energy density averaged over a period,
\begin{equation}\label{e:averaged-E}
\bar{E}=\frac{1}{2\pi}\int_0^{2\pi}d\tau E\,,
 \end{equation}
and in this subsection the bar over a quantity will denote its time average.
Using the results of the $\varepsilon$ expansion, 
Eqs.\ \eqref{e:phi3b1} - \eqref{e:phi3b3},
the time averaged energy density,
 $\bar{{\cal E}}$, up to fourth order in $\varepsilon$ can be written as
\begin{eqnarray}\label{e:averaged-density}
\bar{{\cal E}}&=&\frac{\varepsilon^2}{2\lambda}S^2
-\frac{\varepsilon^4}{216\lambda^3}\biggl[
\lambda S^2(64g_2^2+27\lambda)(S^2+2)
-54\lambda^2(\nabla S)^2
\nonumber\\
&&-SZ(135g_5-378g_2g_4+280g_2^4-36\lambda g_2^2+9\lambda^2)
\biggl] \,.
\end{eqnarray}
For the $\phi^4$ theory the time averaged energy density takes a
much simpler form:
\begin{equation}
\bar{{\cal E}}=\frac{\varepsilon^2}{3}S^2+
\varepsilon^4\left[\frac{1}{6}(\nabla S)^2-\frac{41}{108}S^2(S^2+2)
+\frac{65}{36}SZ\right]S^4 \,.
\end{equation}

Using the above results, the time averaged total energy, $\bar E$,
has the following $\varepsilon$ expansion in $D$ dimension:
\begin{equation}\label{e:Eeps-dep}
\bar{E}=\varepsilon^{2-D}\frac{E_0}{2\lambda}+ \varepsilon^{4-D}E_1\,,
\quad {\rm where}\quad E_0=\int d^{D}\zeta\,S^2\,,
\end{equation}
and $E_1$ denotes the integral
of the $4$-th order term in the energy density,
\eqref{e:averaged-density} in ${\mathbf{R}}^D$.
As one sees from Eq.\ \eqref{e:Eeps-dep} the leading order behaviour
of the time averaged total energy is ${\bar E}\propto\varepsilon^{2-D}$.
This implies that the $\varepsilon$-dependence of $\bar E$ changes essentially
at $D=2$.
In dimensions $D>2$ the total energy {\sl increases} without any bound
for decreasing values of $\varepsilon$.
In $D=2$ $\bar E$ tends to a constant, and for
$D<2$ it goes to zero as $\varepsilon\to 0$.
This also implies that the core energy of a QB in dimensions $D>2$
should exhibit a minimum for some frequency $\omega_{\rm m}$.
In fact, from Eq.\ \eqref{e:Eeps-dep} one immediately finds
\begin{equation}\label{e:omeps-dep}
\omega_{\rm m}^2=1-\varepsilon_{\rm m}^2=
1-\frac{1}{2\lambda}\frac{(D-2)E_0}{(4-D)E_1}\,.
\end{equation}
The above result can only be taken as
an indication of the minimum even if $\varepsilon_{\rm m}\ll1$.
The numerical values of $E_0$ and $E_1$ will be given for the fundamental
solutions in $D=2$ and $D=3$ in
case of spherical symmetry in Section \ref{spherical}.
We note that in $D=1$ $S(\zeta)=\sqrt{2}{\rm sech}(\zeta)$, therefore $E_0=4$,
and the leading order $\varepsilon$ dependence of the energy is given as
$\bar E=2\varepsilon/\lambda+{\cal O}(\varepsilon^3)$.

\subsection{Critical dimension $D=4$}

In the following we
present some simple, although important results concerning
the existence of spatially localized solutions of the master equation, \eqref{e:Sequation}.
First spatially localized solutions of \eqref{e:Sequation} which have a limit
for $|\vec{x}|\to\infty$ should decrease exponentially, since those tending to a constant
exhibit oscillatory behaviour. Next we show
that exponentially localized solutions of Eq.\ \eqref{e:Sequation}
cannot exist for $D\geq4$, implying that small amplitude QB's exist only in dimensions $D<4$.
To see this, consider the following virial identity derived from
equation \eqref{e:Sequation}:
\begin{equation}\label{e:viri1}
\langle(\vec{\nabla}S)^2\rangle+\langle S^2\rangle-\langle S^4\rangle=0\,,
\end{equation}
where $\langle f\rangle:=\int d^Dx f(x)$. Furthermore, another virial identity
can be found
from the scaling transformation ($\vec{x}\to\mu \vec{x}$)
of the action corresponding to \eqref{e:Sequation},
$ \int d^Dx[(\vec{\nabla}S)^2+S^2-S^4/2]$:
\begin{equation}\label{e:viri2}
(D-2)\langle(\vec{\nabla}S)^2\rangle+D\langle S^2\rangle-\frac{D}{2}\langle S^4\rangle=0\,.
\end{equation}
From Eqs.\ \eqref{e:viri1} and \eqref{e:viri2} one immediately finds
\begin{equation}\label{e:crit_dim}
2\langle S^2\rangle+\frac{1}{2}(D-4)\langle S^4\rangle=0\,,
\end{equation}
which equality can only be satisfied if $D<4$.

The absence of small amplitude QB's in more than $3$ spatial dimensions does not
imply per se that oscillons would be also absent if $D\geq4$. As a matter of fact
we have found that small amplitude oscillons exist in dimensions $D\geq4$,
without any apparent limitation on $D$.
These higher dimensional ($D\geq4$) small amplitude oscillons
do not have a well defined, exponentially localized core, and they
cannot be described by small amplitude QB's.
Even the total energy of the core in $D\geq4$ is not well defined.
Interestingly by choosing
suitable initial data, for increasing  energy content
one can achieve that they have very long lifetimes.
Various arguments and numerical studies of spherically symmetric oscillons
in $D$-dimensions by Gleiser \cite{Gleiser04} led him to conjecture
the existence of a critical value of $D$ ($D_c\gtrsim6$) above which no long lived
oscillon states would exist.
The existence of higher dimensional small amplitude oscillons
contradict this conjecture, however, since these contain very large amount of energies
it may have been less obvious to start with such initial data. This might explain why
such objects have been missed.
Also for a fixed amount of energy, the lifetime of oscillons
exhibits a significant decrease for $D>3$.
The results of the recent work \cite{SafTra} show that in $D=5$
the lifetimes becomes as small as a few $100$ (in natural units).
Let us mention here another interesting point.
In Ref.\ \cite{Gleiser04} a very long lived oscillon state has been exhibited in $D=6$.
This object is not in
the class of small amplitude oscillons, and should be understood better.
In any case this provides another example how rich the phase space of time
dependent solutions of a simple non-linear wave equation can be.

\section{Solution of the master equation for spherical symmetry}\label{spherical}

In this Section we consider spherically symmetric configurations, in which
case the PDE's determining the functions $S$, $Z$, etc.\ reduce to ODE's. This
simplifies of course significantly the problem of solving both the master equation and the
associated inhomogeneous ones. We exhibit some numerical solutions of these
equations in $D=2$ and in $D=3$.
We present the solution of the $\varepsilon$ expansion in the $\phi^4$ theory
up to $6$th order.

\begin{figure}[!htbp]
\includegraphics[width=12cm]{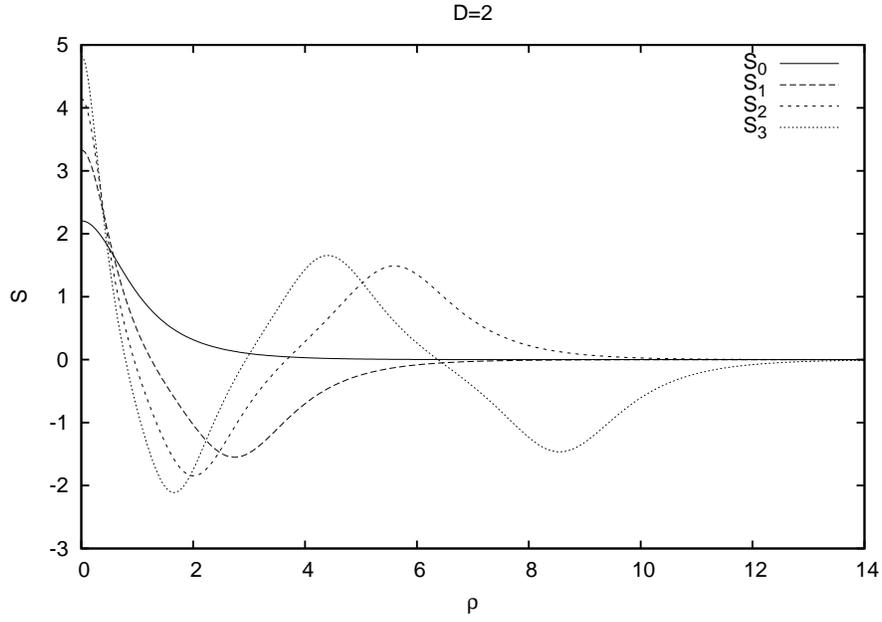}
\caption{\label{f:s2d}
Solutions of the master equation \eqref{e:Seqsph} in $D=2$
with $0, 1, 2$ and $3$ nodes.
}
\end{figure}

\begin{figure}[!htbp]
\includegraphics[width=12cm]{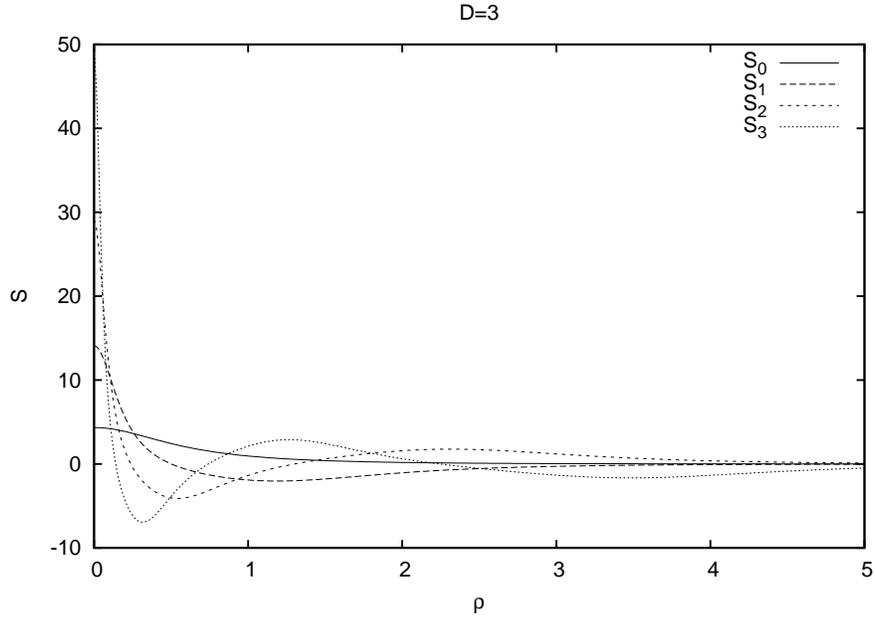}
\caption{\label{f:s3d}
Solutions of the master equation \eqref{e:Seqsph} in $D=3$
dimensions with $0, 1, 2$ nodes.
}
\end{figure}

For spherically symmetric configurations the master equation
\eqref{e:Sequation} takes the form
\begin{equation}
\frac{d^2S}{d\rho^2}+\frac{D-1}{\rho}\,\frac{dS}{d\rho}
-S+S^3=0\,, \label{e:Seqsph}
\end{equation}
where $S$ is a function of the rescaled radial coordinate
$\rho=\varepsilon r$. In $1$ spatial dimension the solution of
\eqref{e:Seqsph} vanishing at infinity is unique, it is given
explicitly by $S=\sqrt 2 \,\mathrm{sech}\, \rho$.
In contradistinction to $D=1$, for higher dimensions, $1<D<4$,
the solution vanishing at $\rho\to\infty$ is not unique.
Our numerical analysis indicates, that for $1<D<4$ there is a family
of localized solutions of Eq.\ \eqref{e:Seqsph} indexed by the
number of zeros (nodes) of $S(\rho)$. On Figures \ref{f:s2d} and
\ref{f:s3d} the first few members of this solution family are
exhibited in dimensions $D=2$ and $D=3$.
As it has been shown in
the previous Section, there are no
solutions of Eq.\ \eqref{e:Seqsph} for $D\geq 4$, which tend to zero for $\rho\to\infty$.
The values of $S$ at the origin $\rho=0$ are tabulated in Table
\ref{t:seq2d3d}.
\begin{table}[htbp]
\begin{tabular}{|c|r @{.}l|r @{.}l|}
\hline
number &
\multicolumn{4}{c|}{$S(\rho=0)$}\\
of nodes&\multicolumn{2}{c}{$D=2$} &
\multicolumn{2}{c|}{$D=3$} \\
\hline\hline
$0$  & $2$&$20620086$  &  $4$&$33738768$
\\ \hline
$1$  & $3$&$33198927$  & $14$&$10358440$
\\ \hline
$2$  & $4$&$15009404$   & $29$&$13121158$
\\ \hline
$3$  & $4$&$82960282$  & $49$&$36070988$
\\ \hline
\end{tabular}
\caption{\label{t:seq2d3d}
Central values of $S$ for two and three dimensions.
}
\end{table}
It can be already expected that the fundamental solution (without nodes), $S_0$,
is physically the most important. Indeed, as we shall show later, oscillons corresponding
to solutions of the master equation with nodes contain more energy and
have significantly smaller
lifetimes than those corresponding to nodeless ones,
and they are also less stable.

\begin{figure}[!htbp]
\includegraphics[width=12cm]{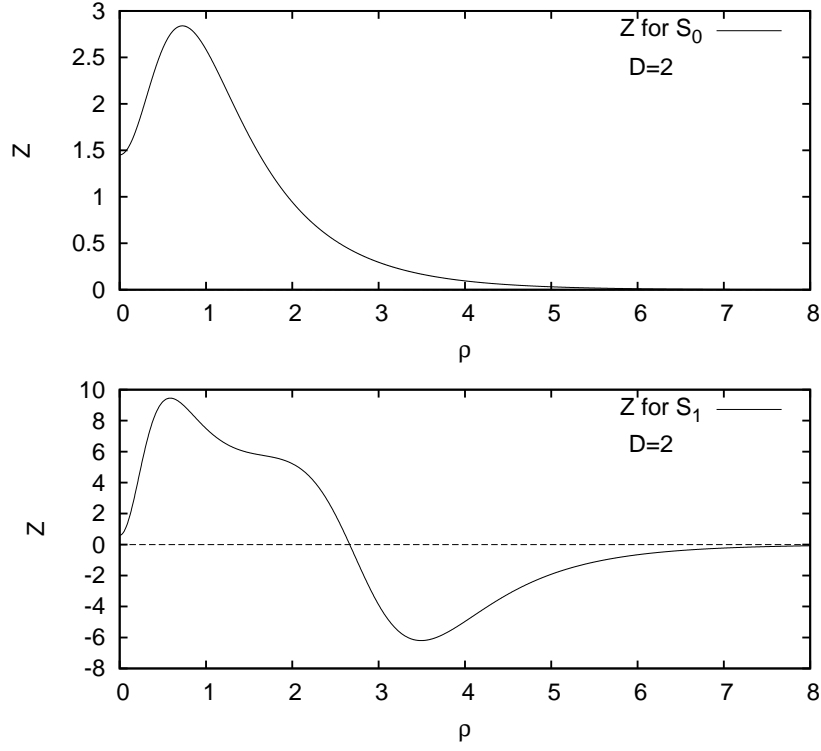}
\caption{\label{f:z2d}
Solutions of Eq.\ \eqref{e:Zsph} for $Z$ corresponding to $S$ without and with $1$
node in two dimensions. The central values are $1.45076$
and $0.6018575$, respectively.
}
\end{figure}

\begin{figure}[!htbp]
\includegraphics[width=12cm]{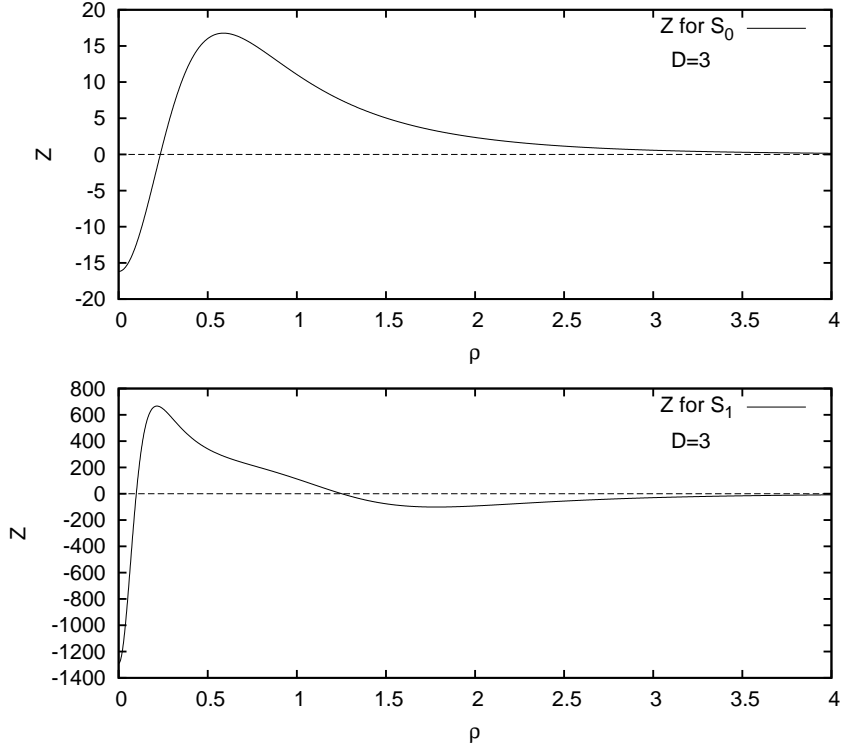}
\caption{\label{f:z3d}
Solutions of Eq.\ \eqref{e:Zsph} in three dimensions. The central values
of $Z$ are $-16.17403$ and $-1290.021$.
}
\end{figure}

In the case of spherical symmetry, equation \eqref{e:Z} determining the third
and fourth order terms in $\varepsilon$ takes the form
\begin{equation}
\frac{d^2Z}{d\rho^2}+\frac{D-1}{\rho}\,\frac{dZ}{d\rho}\,-Z+3S^2Z-S^5=0\,.
\label{e:Zsph}
\end{equation}
Since Eq.\ \eqref{e:Zsph} is linear for the unknown, $Z$, with inhomogeneity
$S^5$ it admits a globally regular solution for any S regular at $\rho=0$ and
vanishing for $\rho\to\infty$.
Some numerical solutions of Eq.\ \eqref{e:Zsph} are plotted on
Figs.\ \ref{f:z2d} and \ref{f:z3d} in two and three dimensions.

\subsection{Solution for the $\phi^4$ theory up to 6th order }

As already discussed, the equation determining the function, $Z$,
is universal for any choice of the potential $U(\phi)$.
Of course the terms in the small amplitude expansion
do depend on the potential $U(\phi)$, and the reconstruction of
$\phi_3$ from $Z$ using Eqs.\ \eqref{e:p3} and \eqref{e:phi3b3}
depends of the values of the coefficients $g_i$.
In the rest of this paper we shall concentrate on the $\phi^4$ theory given
by Eq.\ \eqref{e:phi4th} and provide the results of numerical simulations
only for this case.

The equation determining the fifth order term in the small amplitude
expansion, $\phi_5$, through $p_5$ can be written as
\begin{eqnarray}
&&\frac{d^2Y}{d\rho^2}+\frac{D-1}{\rho}\,\frac{dY}{d\rho}
-Y+3S^2Y+\frac{4225}{64}SZ(3Z-5S^3)\nonumber\\
&&+\frac{53D(D-1)}{\rho^2}S\left(\frac{dS}{d\rho}\right)^2
+\frac{106(D-1)}{\rho}S^2(S^2-1)\frac{dS}{d\rho}
+\frac{8287}{48}S^7
=0\,, \label{e:Yeq}
\end{eqnarray}
where $Y$ is defined by
\begin{equation}
p_5=\frac{\sqrt 2}{9\sqrt 3}\left(
Y-\frac{1235}{32}S^2Z+\frac{1503}{16}Z-24S-\frac{17}{3}S^3
+\frac{11525}{384}S^5\right) \,. \label{e:Ydef}
\end{equation}
The solution of Eq.\ \eqref{e:Yeq} corresponding to the
fundamental solution of the master Eq.\  $S_0$,
has the central value $Y_0(0)=-87.78183$ in two dimensions and $Y_0(0)=60356.38$ in
$D=3$. These actual values themselves have
no physical significance in view of the scaling freedom in the definition
of the function $Y$. Clearly, any constant could have been included in front of the
term $Y$ in Eq.\ \eqref{e:Ydef} without modifying the final
results for the magnitude of the $\phi_i$'s in the $\varepsilon$ expansion.

\begin{figure}[!htbp]
\includegraphics[width=12cm]{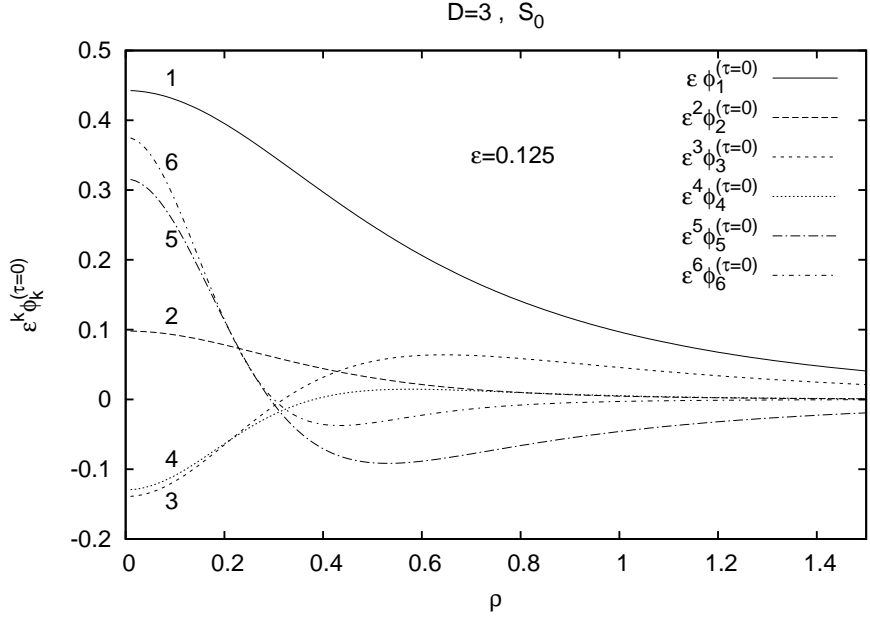}
\caption{\label{f:ph0d3}
Contributions of the various $\varepsilon^k$ order terms corresponding to
$S_0$ to the scalar field, $\phi$, at the moment of time symmetry $\tau=0$ in $D=3$.
The value of $\varepsilon$ has been chosen to be $0.125$, which brings the
contribution of different terms to the same order.
This value of $\varepsilon$ is an obvious upper limit for the range of validity
of our expansion.
}
\end{figure}

\begin{figure}[!htbp]
\includegraphics[width=12cm]{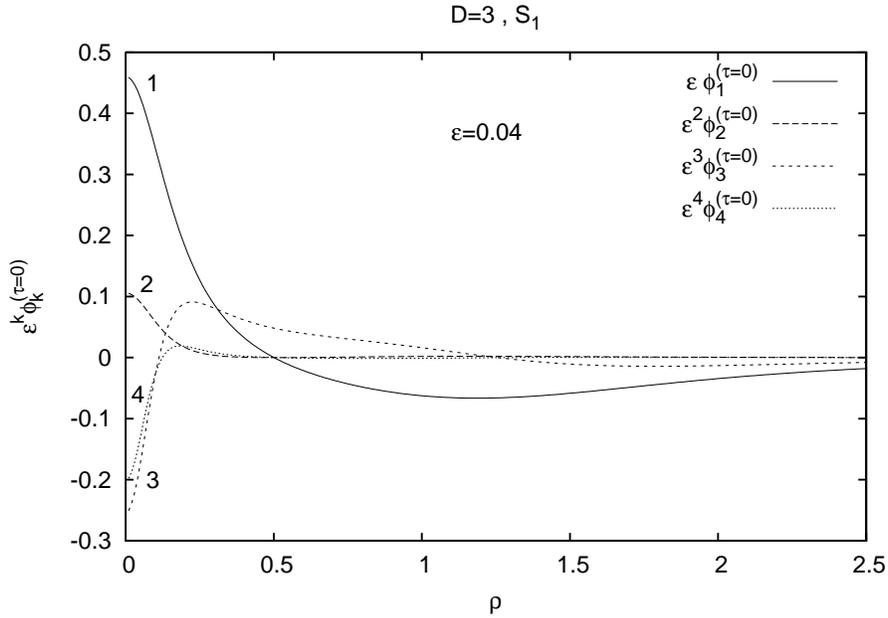}
\caption{\label{f:ph1d3}
$\varepsilon^k \phi_k^{(\tau=0)}$ corresponding to the solution $S_1$, with one node
in the three dimensional case. Because of the sharp increase of the
higher order functions the value of $\varepsilon$ was chosen to be a much smaller as for
$S_0$.
}
\end{figure}

\begin{figure}[!htbp]
\includegraphics[width=12cm]{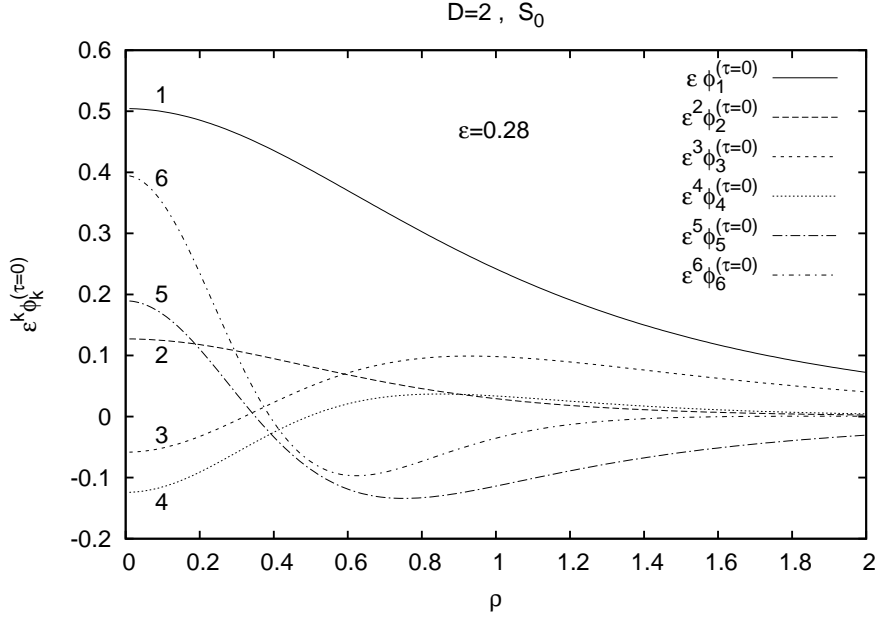}
\caption{\label{f:ph0d2}
$\varepsilon^k \phi_k^{(\tau=0)}$ corresponding to $S_0$ in the two dimensional case.
It can be seen that the value of the
expansion parameter $\varepsilon$ can be chosen here significantly larger
than for $D=3$.
}
\end{figure}

\begin{figure}[!htbp]
\includegraphics[width=12cm]{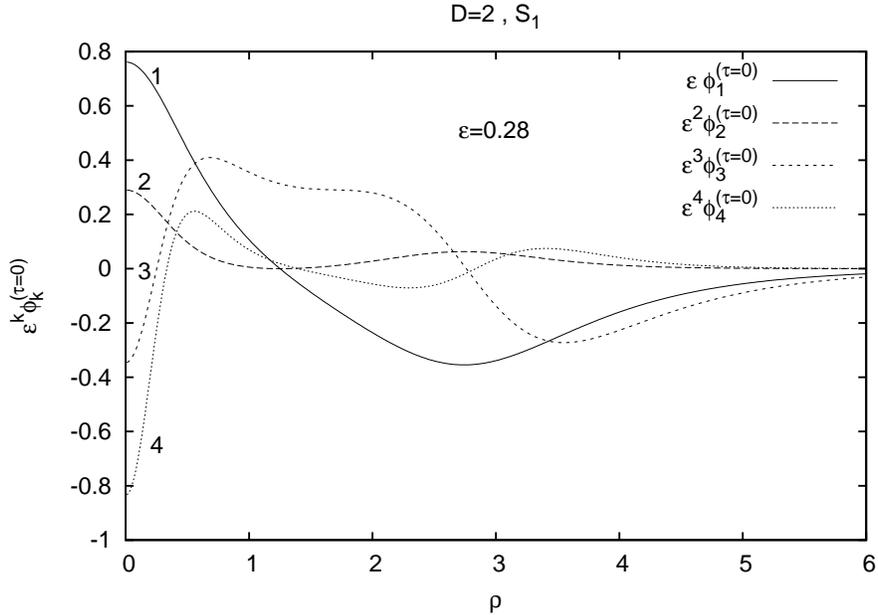}
\caption{\label{f:ph1d2}
$\varepsilon^k \phi_k^{(\tau=0)}$ for $D=2$ and with $S_1$.
The value of $\varepsilon$ is the same as on
Fig. \ref{f:ph0d2}. The contribution of the higher order terms
is bigger than for $S_0$, nevertheless it
is smaller than for $D=3$.
}
\end{figure}

In the following we list the values of the terms of the small amplitude expansion,
$\phi_i$, up to order six in the $\phi^4$ theory,
at the moment of time reflection symmetry, $\tau=0$:
\begin{eqnarray}
\phi_1^{(\tau=0)}&=&\sqrt{\frac{2}{3}}\, S\label{e:ph01}\\
\phi_2^{(\tau=0)}&=&\frac{1}{3}S^2\\
\phi_3^{(\tau=0)}&=&\frac{1}{9}\sqrt{\frac{2}{3}}\left(
\frac{195}{8}Z-8S-\frac{35}{8}S^3\right)\\
\phi_4^{(\tau=0)}&=&\frac{1}{9}\left[
\frac{65}{4}SZ+10\left(\frac{dS}{d\rho}\right)^2
+\frac{8}{3}S^2-\frac{125}{12}S^4
\right]\\
\phi_5^{(\tau=0)}&=&\frac{1}{9}\sqrt{\frac{2}{3}}\Biggl[
Y-\frac{2275}{64}S^2Z+\frac{1503}{16}Z
-\frac{15}{32}S\left(\frac{dS}{d\rho}\right)^2
-24S-\frac{595}{96}S^3+\frac{11285}{384}S^5
\biggr]\\
\phi_6^{(\tau=0)}&=&\frac{2}{27}\Biggl[
SY+\frac{325}{4}\frac{dS}{d\rho}\frac{dZ}{d\rho}
+\frac{4225}{128}Z^2-\frac{8125}{48}S^3Z+\frac{6589}{48}SZ
-\frac{9223}{32}S^2\left(\frac{dS}{d\rho}\right)^2
+\frac{88}{3}\left(\frac{dS}{d\rho}\right)^2\nonumber\\
&&+\frac{26D(D-1)}{\rho^2}\left(\frac{dS}{d\rho}\right)^2
+\frac{52(D-1)}{\rho}S(S^2-1)\frac{dS}{d\rho}
+\frac{92}{9}S^2-\frac{35417}{288}S^4
+\frac{21467}{144}S^6
\biggr]\,.\label{e:ph06}
\end{eqnarray}
These expressions are presented because they are needed to
provide good initial data for numerical time evolution simulations.
They will be actually used in the next Section \ref{s:evolution}.
Since the range of $\phi_k^{(\tau=0)}$ increases very much with $k$, we
depict the product $\varepsilon^k \phi_k^{(\tau=0)}$ for some chosen values of
$\varepsilon$ on
Figs. \ref{f:ph0d3}, \ref{f:ph1d3}, \ref{f:ph0d2} and \ref{f:ph1d2}.
On these four figures $S_0$ and $S_1$ are depicted in spatial dimensions
$D=2$ and $D=3$.
In a given order in the expansion, the value of $\varepsilon$ which brings the contribution of lower order terms approximately to
the same order of magnitude will be used as an upper estimate for the range of
$\varepsilon$, below which our expansion can still be expected to yield an acceptable
approximation.
Generally speaking, in an asymptotic expansion
for a given (small) value of the expansion parameter one can only sum terms up to such
an order until which all terms decrease.
A simple comparison of a QB corresponding to the fundamental solution, $S_0$,
with another one corresponding to a solution with a single node,
$S_1$, (compare Figures \ref{f:ph0d3} and \ref{f:ph1d3}),
makes one to guess that oscillons containing fundamental QB's
are likely to have better stability properties and longer lifetimes, than those
containing QB's based on solutions with nodes, at least in $D=3$.
Our numerical simulations show that this is indeed the case (see Section
\ref{s:evolution}).
In two dimensions the difference between the longevity and stability properties
of oscillons containing QB's corresponding to solutions of the $S$ equation \eqref{e:Sequation}
with nodes is much less pronounced than in three spatial dimensions.
It is also apparent that the $\varepsilon$ expansion is valid for significantly
larger values of $\varepsilon$ in the two dimensional case than in the three
dimensional one.

\begin{figure}[!htbp]
\includegraphics[width=12cm]{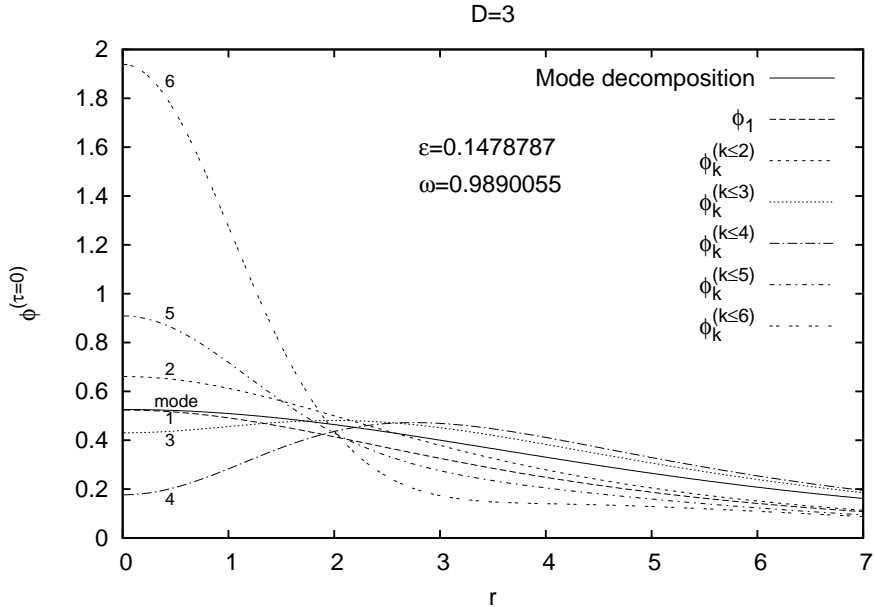}
\caption{\label{f:cmp398b} Comparison of the value of the field at
$\tau=0$ obtained by the $\varepsilon$ expansion to a very precise value
obtained by a high order Fourier mode decomposition for
$\omega=0.9890055$, i.e. for $\varepsilon=0.1478787$ in $D=3$.
For such a large value of $\varepsilon$ the first order
approximation gives a remarkably good estimate in the neighbourhood of
the central region.
The contributions of the higher order approximations
make this agreement increasingly worse in the neighbourhood of
the origin.
Farther away from the center, however, the third order
approximation gives the best, although not a very precise, result. }
\end{figure}

\begin{figure}[!htbp]
\includegraphics[width=12cm]{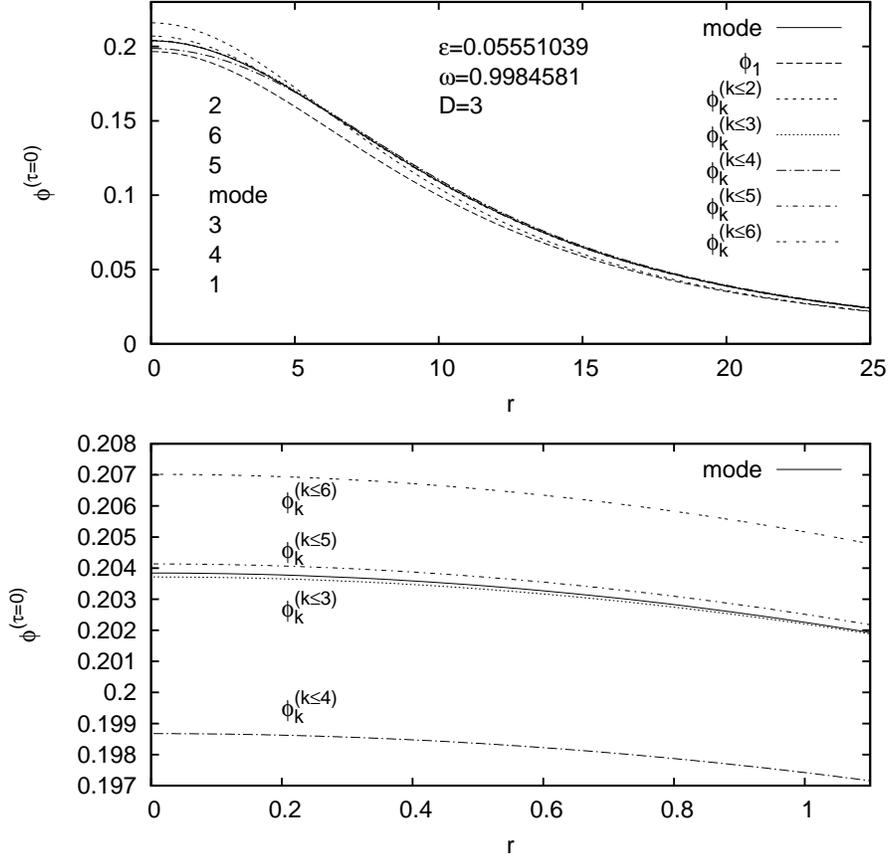}
\caption{\label{f:cmp41}
Comparison of various order $\varepsilon$ expansions of $\phi$ at $\tau=0$ to the
value obtained by Fourier mode decomposition for $\omega=0.9984581$ i.e. for
$\varepsilon=0.05551039$ in the $D=3$ case.
}
\end{figure}

In order to get an independent check on the validity of the
small amplitude expansion, we
have compared some QB's up to order $6$, with time periodic QB's obtained
previously by solving the NLWE \eqref{e:evol} directly by Fourier mode decomposition
\cite{FFGR}.
On Figures \ref{f:cmp398b} and \ref{f:cmp41} we depict $\phi$ computed to
various orders in the $\varepsilon$-expansion, and also the QB obtained
in Ref.\ \cite{FFGR} by Fourier mode decomposition using very precise spectral methods
provided by the LORENE library \cite{spectral}.
The chosen frequencies correspond
to two states investigated in detail in \cite{FFGR}. Note that in Ref.\ \cite{FFGR}
the interaction potential had a different scale, and the resulting threshold frequency
was $\sqrt{2}$ as opposed to the value $1$ in the present paper.
The periodic quasi-breather solutions chosen from Ref.\ \cite{FFGR}
have frequencies $\tilde\omega=1.412033$ and $\tilde\omega=1.398665$ (see Figs
5. and 19. in \cite{FFGR}). In the present conventions these values correspond to
frequencies $\omega=0.9984581$ and $\omega=0.9890055$. The corresponding values
of $\varepsilon$ are: $\varepsilon=0.05551039$ and $\varepsilon=0.1478787$.
It can be seen on Figures \ref{f:cmp398b}, \ref{f:cmp41}
that for higher values of $\varepsilon$, only
the leading term or eventually the first two orders in the $\varepsilon$
expansion give meaningful results. The relatively big error in these approximation
cannot be decreased because of the asymptotic nature of the expansion.
As $\varepsilon$ gets smaller and the frequency $\omega$ gets closer
to the basis frequency $1$, more and more higher order terms in the $\varepsilon$
expansion can be used, and then the error also decreases significantly.
It can be seen on Fig.\ \ref{f:cmp41} that for the smaller value
$\varepsilon$ although the second, third and fifth order
expansion gives an improvement on the lower order values, the fourth
and sixth order expansion turns out to be less precise than the third
and fifth order expressions. We think that this is related to the fact
that the signature of the various order contributions changes in
pairs, i.e. $\phi_{k}^{(\tau=0)}$ is positive at the center $r=0$ for
$k=1,2,5,6...$ and negative for $k=3,4,7,8...$, as can be seen on
Fig.\ \ref{f:ph0d3}. This alternating improving and not improving
behaviour for odd and even orders happens only for intermediate values
of $\varepsilon$. For even smaller $\varepsilon$, e.g.\ for
$\varepsilon=0.01$, the error decreases monotonically when increasing
the order of the expansion.

Let us now come back to the $\varepsilon$-expansion of the energy
\eqref{e:Eeps-dep} computed in subsection \ref{ss:energy}.
From our results it is not difficult to calculate numerically $E_0$ and $E_1$ in $D=2$
and $D=3$.
We find that the first two terms of the time averaged energy
in the $\varepsilon$-expansion are given as:
\begin{equation} \label{e:sph-omeps-dep}
 {\bar E}\approx3.9003+26.9618\varepsilon^2\,,\quad {\rm for\ }D=2\,,\quad
  {\bar E}\approx6.29908/\varepsilon+264.262\varepsilon\,,\quad {\rm for\ }D=3\,.
\end{equation}
This simple estimate Eq.\ \eqref{e:sph-omeps-dep} gives
for the minimal value of $\varepsilon_{\rm m}\approx0.15428$, which is
unfortunately already too large to be trusted.
Nevertheless it can still be accepted as the indication that
such a minimal value, $\omega_{\rm m}$ exists.
For spherically
symmetric oscillons in $D=3$ it has been found that $\omega_{\rm m}\approx0.9659$
\cite{FFGR}. This value of $\omega_{\rm m}$ corresponds to
$\varepsilon_{\rm m}\approx0.2588$ which value is way too large for us.

\section{Time evolution}\label{s:evolution}

The precision and applicability of the $\varepsilon$ expansion can be
checked by using the field value obtained by the expansion as initial
data for a numerical time evolution code applied for our spherically
symmetric scalar field system.  The field value given by
Eqs. \eqref{e:ph01} - \eqref{e:ph06} and \eqref{e:sumphi} at $\tau=t=0$
has been used as initial data for various $\varepsilon$ values in
$D=2$ and $D=3$ spatial dimensions.

The applied numerical evolution code is a slightly modified version of
the fourth order method of line code used in \cite{FFGR} for studying
oscillons and developed in
\cite{FR} for the study of spherically symmetric
magnetic monopole configurations.
The spatial grid is chosen to be uniform in the compactified
radial coordinate $R$ defined by
\begin{equation}
r=\frac{2R}{\kappa(1-R^2)} \,,
\end{equation}
where $\kappa$ is a constant which may be chosen differently though for
each choice of initial data. The whole range
$0\leq r<\infty$ of the physical radial coordinate $r$ is mapped to
the interval $0\leq R<1$, avoiding the need for explicitly describing
boundary conditions at some large but finite radius.  Since the
characteristic size of the obtained oscillon states is inversely
proportional to $\varepsilon$ we chose $\kappa$ to be proportional to
$\varepsilon$, keeping the oscillon occupying approximately the same
region of the $R$ coordinate range. For the actual calculations we
used $\kappa=5\varepsilon$.

Using the radial coordinate $R$ the field equation \eqref{e:evol} takes the
form
\begin{equation}\label{e:phitR}
\phi_{,tt}=
\frac{\kappa^2(1-R^2)^3}{2(1+R^2)}\left[
\frac{(1-R^2)}{2(1+R^2)}
\phi_{,RR}
-\frac{R(3+R^2)}{(1+R^2)^2}
\phi_{,R}
+\frac{(D-1)}{2R}
\phi_{,R}
\right]
-U'(\phi) \,.
\end{equation}
Introducing the new variables
\begin{eqnarray}
\phi_{t}&=&\phi_{,t}\label{e:ev2}\\
\phi_{R}&=&\phi_{,R}\label{e:constr}
\end{eqnarray}
the problem can be interpreted as a system of first order differential equations
comprising \eqref{e:ev2}, \eqref{e:phitR}, and $\phi_{R,t}=\phi_{t,R}$ for
the three variables $\phi$, $\phi_{t}$ and $\phi_{R}$.
If equation \eqref{e:constr}
holds at $t=0$ it is preserved by the evolution equations, thereby it can be
considered as a constraint. The third term inside the bracket on the right hand
side of \eqref{e:phitR} cannot be directly evaluated numerically at the center
$R=0$. At the grid point corresponding to the center this term is calculated
using the identity
\begin{equation}
\lim_{r\to0}\frac{\phi_{,R}}{R}=\lim_{r\to0}\phi_{,RR} \,.
\end{equation}
Since the evolution of the initial data provided by the $\varepsilon$ expansion
show different characteristics in two and three dimensions we discuss these
cases in different subsections.

\subsection{$D=3$}

In the three dimensional case, 
there are two different types of oscillons, 
a stable and an unstable type. 
For the frequency range $\omega<\omega_{\rm c}\approx0.967$, 
(i.e. for $\varepsilon>0.255$), oscillons are
essentially stable. They slowly radiate energy while 
their frequency, $\omega(t)$, increases towards a critical frequency
$\omega_{\rm c}$.
When they reach the critical frequency
these oscillons quickly disintegrate.
Oscillons with $\omega>\omega_{\rm c}$ have one
unstable decay mode, 
which can be suppressed by fine tuning the initial
data. Close to the critical value of the parameter in the initial data
there can be two types of decay mechanisms. One with a uniform
outwards motion of the energy, and another through a temporary
collapse to a small central region (see Fig.\ 4 of \cite{FFGR}). In
order to find such oscillons a very precise fine tuning of the initial
data is necessary. For example in Ref.\ \cite{FFGR} this fine tuning
corresponded to the classical bisection procedure between two values
of a suitable parameter in the initial data yielding the two different
decay modes. This way one obtains very long living oscillon states.
The frequency of these unstable oscillons decreases slowly towards 
$\omega(t)\to\omega_{\rm c}$.

Remarkably the energy of QB's as a function of their frequency exhibits a minimum 
at $\omega\approx\omega_{\rm c}$, (see Fig. 3 in \cite{Watkins},
Fig.\ 17 of \cite{FFGR} and Fig.\ 4 of \cite{SafTra}). 
Therefore it is natural to assume that the two types of oscillons
are also distinguished by the same
behaviour of their energy as function of their oscillon frequency.

Unfortunately, for values of $\varepsilon\gtrsim0.25$
initial data obtained by the $\varepsilon$ expansion are well outside
the domain of validity of the expansion. Using such initial data
gives decaying states which are unrelated to the stable oscillons with the
intended frequency. 
On the other hand the $\varepsilon$ expansion yields good initial data
for small amplitude unstable oscillons. 
As a matter of fact the
initial data obtained this way makes
a fine-tuning procedure of the initial data unnecessary.
For sufficiently small values of
$\varepsilon\lesssim 0.1$ the first few terms (at least to order
${\cal O}(\varepsilon^3)$) of the series expansion \eqref{e:ph01} -
\eqref{e:ph06} already yield sufficiently good initial data which
evolve directly into long living oscillon states.

\begin{figure}[!htbp]
\includegraphics[width=12cm]{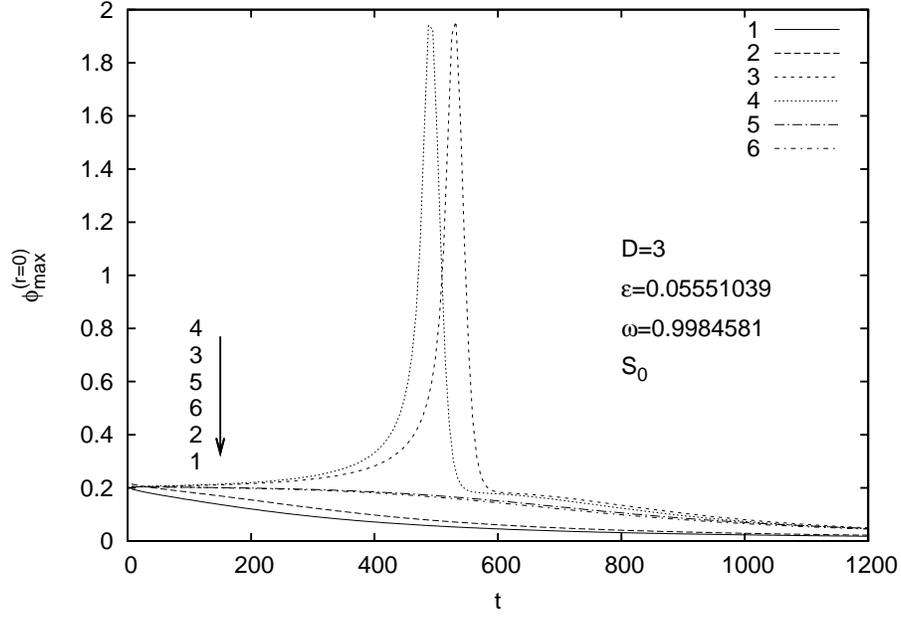}
\caption{\label{f:ev3d055}
Time evolution of the initial data obtained by the $\varepsilon$ expansion
method up to order six for $\varepsilon=0.05551039$ in three spatial
dimensions.
}
\end{figure}

\begin{figure}[!htbp]
\includegraphics[width=12cm]{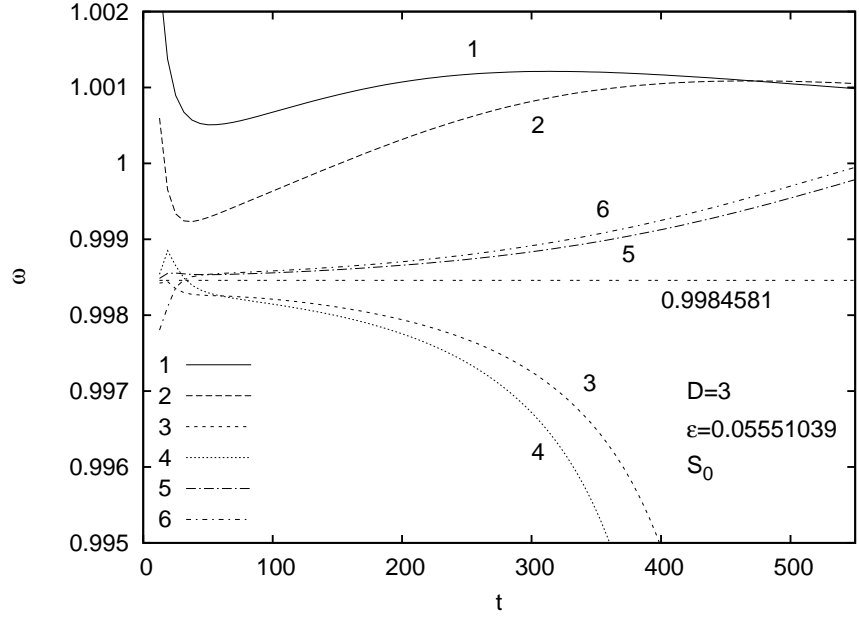}
\caption{\label{f:ev3d055fr}
Oscillation frequency as a function of time for the states shown on Fig.
\ref{f:ev3d055}.
}
\end{figure}

\begin{figure}[!htbp]
\includegraphics[width=12cm]{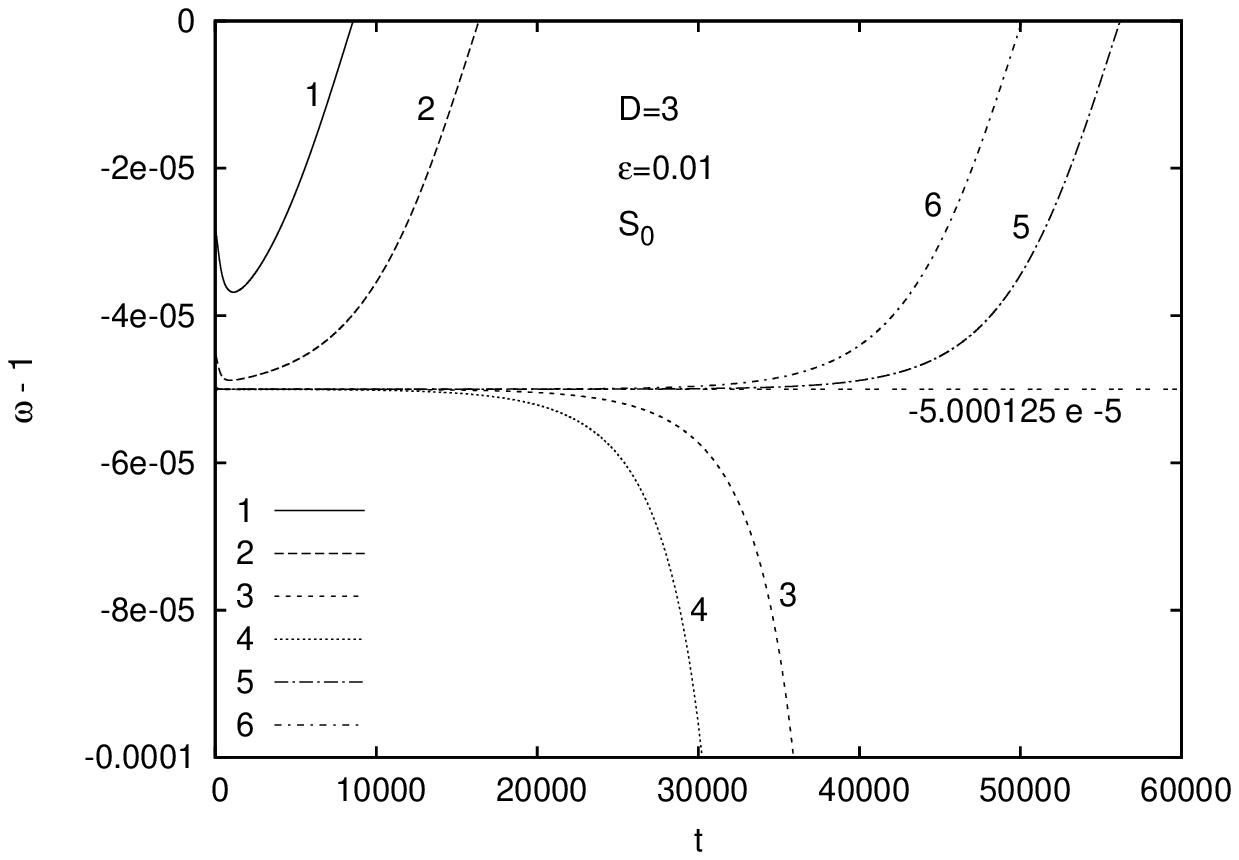}
\caption{\label{f:ev3d01fr}
Time evolution of the frequency of the oscillations of $\phi$ evolving from
initial data with $\varepsilon=0.01$ in the $D=3$ case.
Since the frequency is very close to one, the value $\omega-1$ is plotted
instead of $\omega$.
}
\end{figure}

\begin{figure}[!htbp]
\includegraphics[width=12cm]{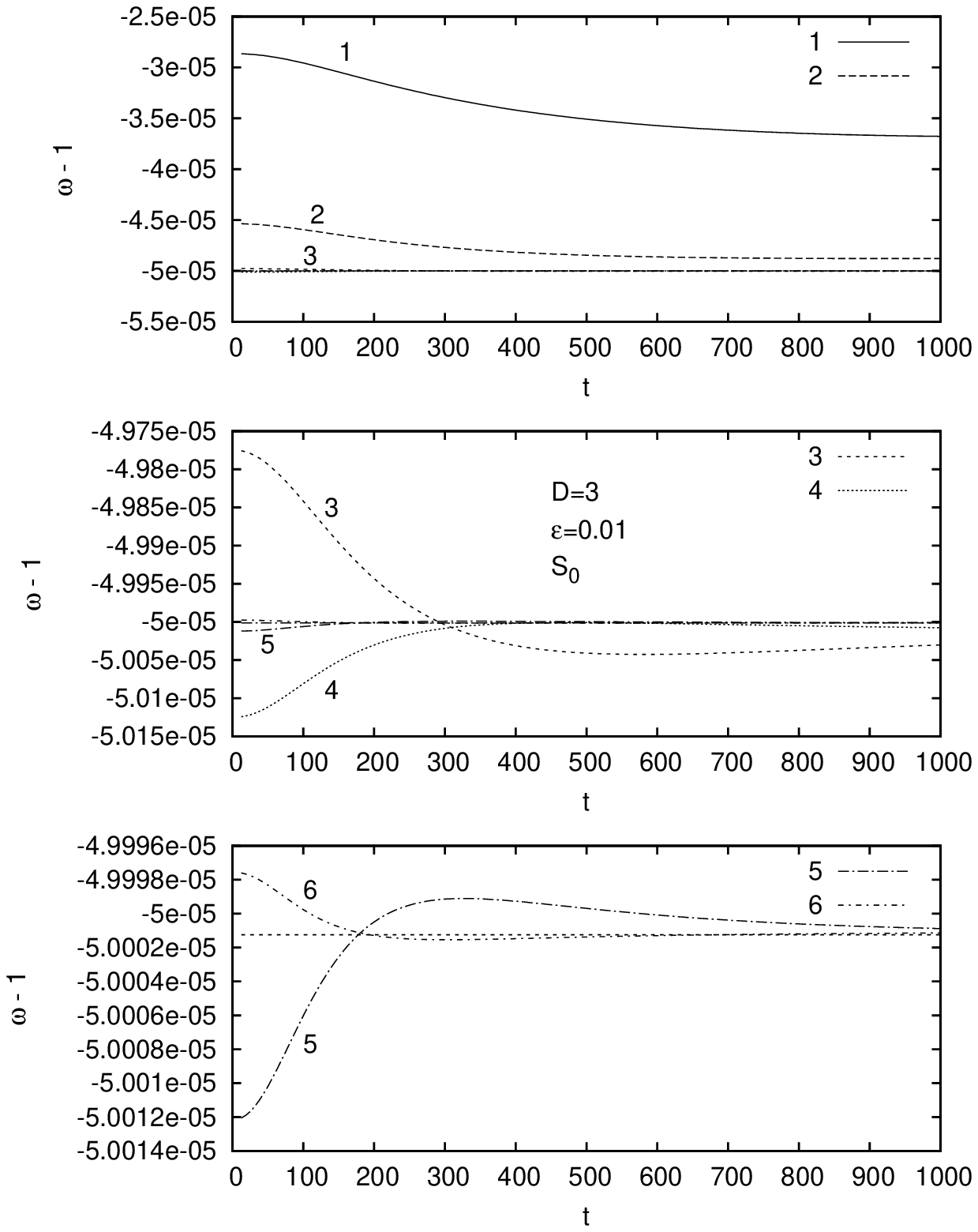}
\caption{\label{f:ev3d01b}
Beginning of the time evolution simulations shown on Fig.\ \ref{f:ev3d01fr}.
For this shorter time interval the difference from the expected solution with
$\omega=0.99994999875$ decreases monotonically with the order of the
$\varepsilon$ approximation.
}
\end{figure}

For the three dimensional case we present the results of the time
evolution of the initial data obtained up to order six by the
$\varepsilon$ expansion method for two choices of $\varepsilon$.
First we consider initial data obtained from the basic solution $S_0$
of \eqref{e:Sequation} without nodes. As we will see shortly, those
with nodes provide initial data that leads to states of significantly
shorter lifetimes.  Figure \ref{f:ev3d055} shows the upper envelope of
the central value of $\phi$ for the initial data belonging to
$\varepsilon=0.05551039$.
The decay method changes with the order of the initial data.  The
amplitude peak on the evolution of the third and fourth order initial
data reflects the collapsing decay mode of the oscillon state.
Although, in general, lifetimes get longer for higher order
approximations, this increase is not monotonic.  In accordance with
Fig.\ \ref{f:cmp41}, approximations of order $4$ and $6$ do not bring
any improvement on the the functions of order $3$ and $5$.  The time
evolution of the frequency of the oscillations is plotted on Fig.\
\ref{f:ev3d055fr}.
It can be seen that for this relatively high value of $\varepsilon$
the first and second order approximation yields a shorter living state
quite different from the expected oscillon state with frequency
$\omega=0.9984581$.  On Figure \ref{f:ev3d01fr} the evolution of a
higher frequency initial data with $\varepsilon=0.01$ is shown.
It can be seen that initial data with smaller $\varepsilon$ provide evolutions
with significantly longer lifetimes.
Fig.\ \ref{f:ev3d01b} shows the initial stage of the evolution.
It can be seen that although for a short time the error of the
solution decreases monotonically with the order of the $\varepsilon$
expansion, for longer time intervals the fourth and sixth order
approximations do not improve on the previous order expansions.

\clearpage

\begin{figure}[!htbp]
\includegraphics[width=12cm]{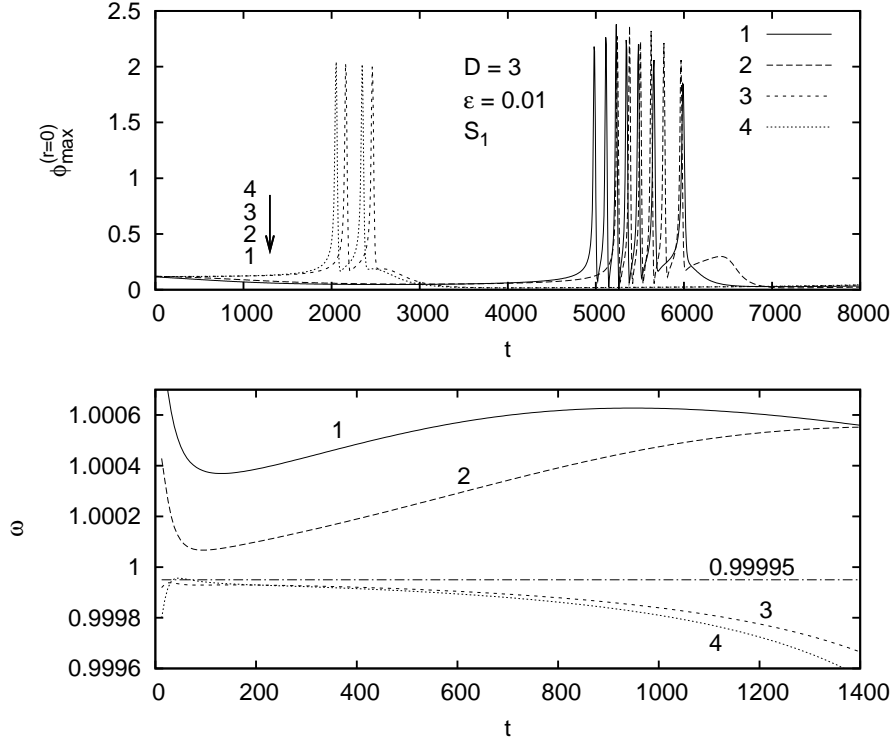}
\caption{\label{f:ev3dz1}
Time evolution of the initial data obtained by the $\varepsilon$ expansion up
to order four using the solution $S_1$ of \eqref{e:Sequation}. Space is three
dimensional and $\varepsilon=0.01$. The top graph shows the upper envelope of
$\phi$ at $r=0$, while the lower graph plots the evolution of the oscillation
frequency. It can be seen that the first and second order initial data
still evolve to states very different from the one with the expected frequency.
}
\end{figure}

Figure \ref{f:ev3dz1} shows the evolution of various order initial
data with the same $\varepsilon$ as on the previous two figures but
obtained by using the solution $S_1$ of \eqref{e:Sequation} with one
node.
These are localized, although big size, high energy states.  The time
dependence of these states is rather complex; there is an interior
part ($r\lesssim 50$) which after an initial time interval ($\approx
1200$) shows a complicated time dependence, with large amplitude
variations.  The time dependence of these states is close to being
periodic for recurrent time intervals. In the initial stages these
initial data still evolve close to a periodic configuration with the
expected frequency, although it can stay near this state much shorter
time than the evolution obtained using $S_0$. Decreasing the value of
$\varepsilon$ increases the lifetime of these states as well, but they
still remain less stable and shorter living than the basic states
obtained by using $S_0$.

\begin{figure}[!htbp]
\includegraphics[width=12cm]{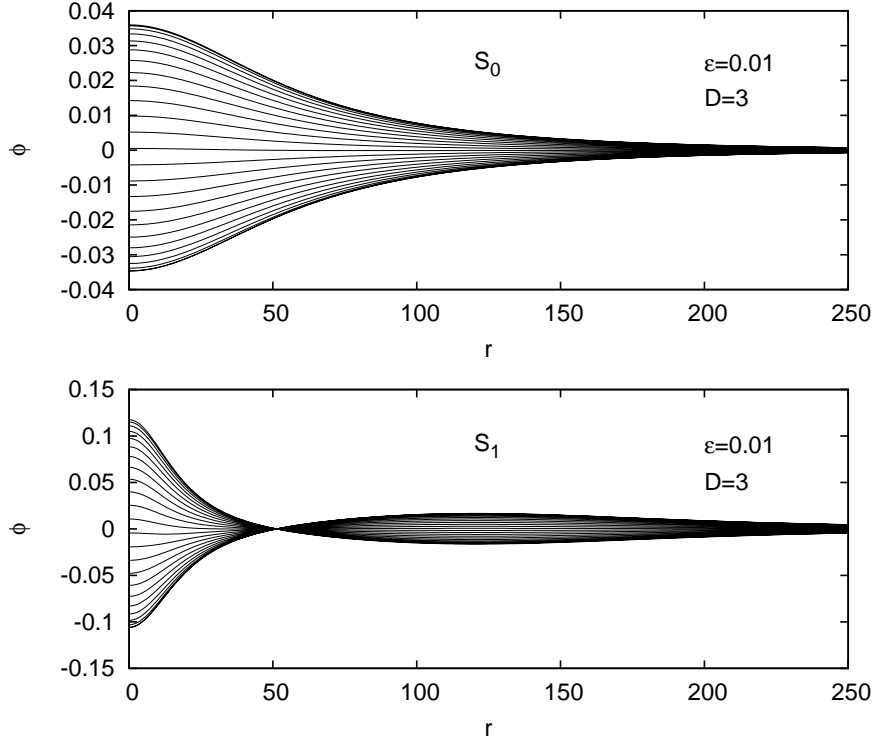}
\caption{\label{f:space}
Time evolution of $\phi$ during a half period of oscillation, plotted as a
function of the radial distance $r$. The upper plot corresponds to initial data
obtained from $S_0$, the lower from $S_1$.
The absolute value of $\phi$ at $r=51.3$ remains below $10^{-5}$.
Both plots contain lines corresponding to uniform time steps between the first
maximum and minimum after $t=300$, although the plot would remain very similar
in a large time interval. The initial data in the $S_0$ case was generated by a
sixth order $\varepsilon$ expansion, while a third order expansion was used in
the $S_1$ case.
}
\end{figure}

\begin{figure}[!htbp]
\includegraphics[width=12cm]{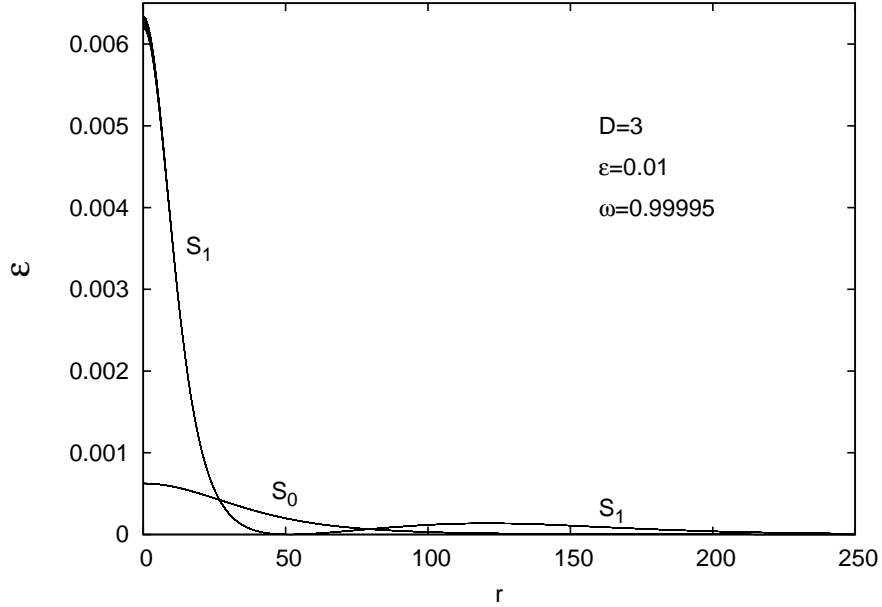}
\caption{\label{f:spacen} Energy density ${\cal{E}}$ of oscillon
  configurations evolved from $S_0$ and $S_1$ initial data, using
  $\varepsilon=0.01$.  The energy density is plotted on the same
  uniformly placed moments of time as on Fig. \ref{f:space}. The
  minimum of ${\cal{E}}$ in the $S_1$ case remains below $10^{-7}$
  near $r=51.3$.  }
\end{figure}

On Figure \ref{f:space} the value of $\phi$ as a function of the radial
coordinate $r$ for subsequent time slices is plotted during a half period of
oscillation. For comparison, the corresponding configuration obtained from
initial data generated with $S_0$ is also presented.
In case of $S_1$ initial data the value of $\phi$ remains very close to
zero at $r=51.3$. The energy density ${\cal{E}}$ also remains very small at this
radius, as can be seen on  Fig.\ \ref{f:spacen}.
The total energy of the $S_0$ configuration is $E=632.536$, while the $S_1$
configuration contains significantly more energy, $E=4061.88$.

\clearpage

\subsection{$D=2$}

In two spatial dimensions there seems to be a single type of oscillon,
which is stable. Once formed, oscillons in $D=2$ are not observed to disintegrate.
Their energy is a monotonically
decreasing function of the frequency. This explains the observation that all oscillon
states are stable. When they slowly emit energy by
radiation they gradually evolve through oscillon states with increasing
frequency, $\omega(t)\to1$. Oscillons evolve from a wide range of initial
data, by shedding most of the surplus energy quickly during an initial
state. However, in general, a slow periodic change can be seen on the
amplitude and on the frequency of the oscillations, indicating a
breathing type oscillation of the oscillon as a whole. The amplitude
of this low frequency ringing depends on how closely the initial data
approaches a given pure oscillon state.

\begin{figure}[!htbp]
\includegraphics[width=12cm]{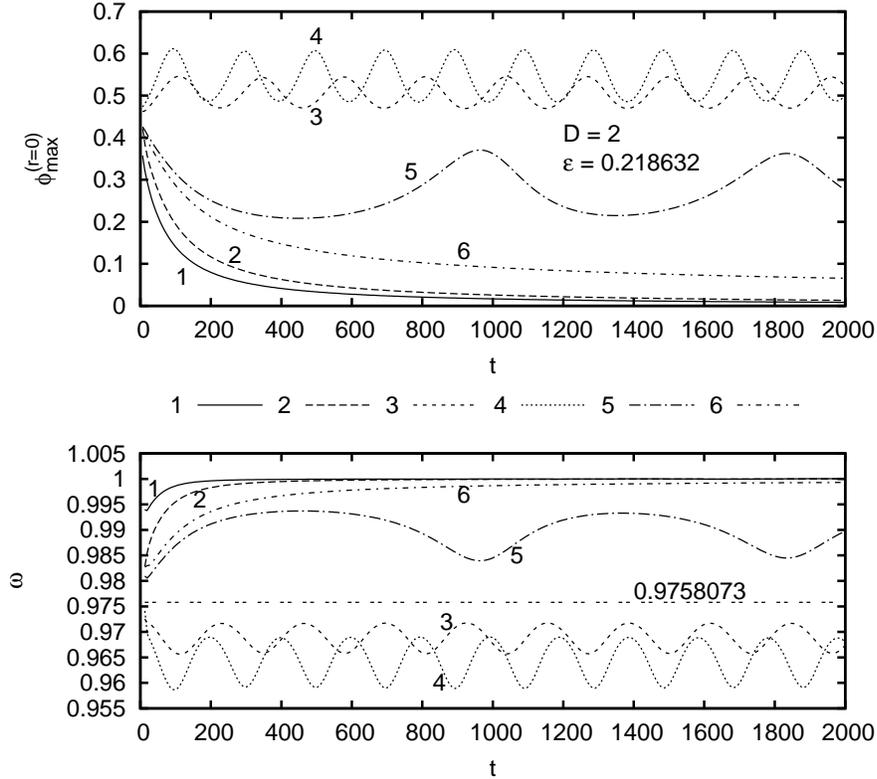}
\caption{\label{f:ev2d21}
Time evolution of the initial data produced by various order $\varepsilon$
expansions for $\varepsilon=0.218632$ in the two dimensional case. The top
graph shows the upper envelope of $\phi$ at $r=0$, while the bottom graph the
time dependence of the frequency. The evolution of the third and fourth order
initial data clearly corresponds to some oscillon state deformed by a low
frequency ringing mode. The evolution of the fifth order initial data
corresponds to a similar state with extreme high amplitude ringing. The
first, second and sixth order initial data evolve into decaying modes,
indicating a large error at these orders of the $\varepsilon$ expansion.
}
\end{figure}

Since, similarly to the $D=3$ case, evolutions from initial data
obtained using solutions $S$ of \eqref{e:Sequation} with nodes produce
less stable and shorter living states, in the following we present
only numerical simulations corresponding to the nodeless solution
$S_0$.  In two dimensions the validity domain of the expansion extends
to significantly higher values of $\varepsilon$. On Fig.\
\ref{f:ev2d21} we present time evolution results for
$\varepsilon=0.218632$ corresponding to $\omega=0.9758073$.
It can be seen that at this high $\varepsilon$ value the approximation improves
up to order tree in the expansion, and then starts to deteriorate in line with
the asymptotic nature of the expansion.

\begin{figure}[!htbp]
\includegraphics[width=12cm]{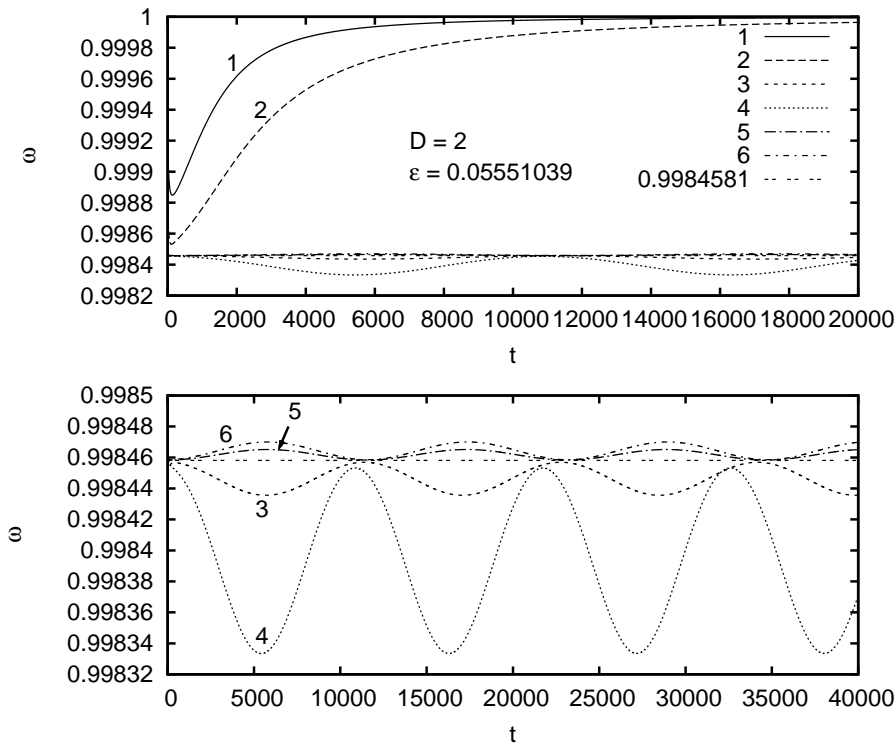}
\caption{\label{f:ev2d055}
The frequency as a function of time for the evolution of initial data with
$\varepsilon=0.05551039$ in case of $D=2$.
}
\end{figure}

On Fig.\ \ref{f:ev2d055} the $\varepsilon=0.05551039$ case is presented.
This value of $\varepsilon$ with the corresponding frequency
$\omega=0.9984581$ has been also studied in $D=3$ dimensions.
In this intermediate frequency case the first and second order initial data
still lead to decaying evolutions.
Higher order expansions tend to give improving approximations of an
oscillon state ringing with a low frequency.
Increasing the order of the expansion gives smaller amplitude ringings, thereby
approaching a pure, very closely periodic, oscillon state.
Similarly to the three dimensional case, initial data of order four and six
yield larger error than order three and five, which, however, does not
change the overall improving tendency of the initial data.

The average amplitude and average frequency of the presented states
show extremely little change even for much longer time periods than
the ones presented on the figures.  The amplitude of the low frequency
ringing decreases very slowly too.  To determine the rate of change of
the amplitude and the corresponding slow energy loss by radiation
would require very high resolution numerical runs requiring excessive
processor time.

\section{Conclusions}\label{s:conc}
Small amplitude oscillons represent an important subset
of time dependent long-living lumps.
We have shown that they can be very well approximated by an asymptotic
series of localized, time-periodic breather-like objects (quasi-breathers).
We have developed a general framework to derive the asymptotic series
expansion of small amplitude quasi-breathers, in $D$ spatial dimensions
in general scalar theories.
We have derived a 2nd order elliptic PDE with a cubic non-linearity,
universal for scalar models, which determines these quasi-breathers.
Our numerical investigations in $\phi^4$-theories show that the small amplitude
quasi-breathers obtained by the asymptotic expansion, provide excellent
initial data for long-living oscillons in $D=2$ and $D=3$.
We have found that small amplitude QB's do not exist for $D\geq4$.

\section{Acknowledgments}

This research has been supported by OTKA Grants No. K61636,
NI68228.  G. F. would also like to thank the Grid Application
Support Centre at SZTAKI.

\end{document}